 \documentclass[10pt,preprint]{aastex}  
 \usepackage{natbib}
\def\ang{\AA}
\def\arcsec{\hbox{$^{\prime\prime}$}}

\def\gapprox{\lower.4ex\hbox{$\;\buildrel >\over{\scriptstyle\sim}\;$}}
\def\lapprox{\lower.4ex\hbox{$\;\buildrel <\over{\scriptstyle\sim}\;$}}

\shortauthors{ASCHWANDEN & ET AL 2016}
\shorttitle{Width Distribution of Coronal Loops}

\begin{document}

\title{         The Width Distribution of Loops and Strands in the Solar Corona
		-- Are we Hitting Rock Bottom ? }

\author{        Markus J. Aschwanden$^1$}

\affil{		$^1)$ Lockheed Martin, 
		Solar and Astrophysics Laboratory, 
                Org. A021S, Bldg.~252, 3251 Hanover St.,
                Palo Alto, CA 94304, USA;
                e-mail: aschwanden@lmsal.com}

\and

\author{	Hardi Peter$^2$		}

\affil{		$^2)$ Max Planck Institute for Solar System Research, 
		Justus-von-Liebig Weg 3, 37077 G\"ottingen, Germany; 
		e-mail: peter@mps.mpg.de}

\begin{abstract}
In this study we analyze {\sl Atmospheric Imaging Assembly (AIA)} and Hi-C
images in order to investigate absolute limits for the finest loop strands.
We develop a model of the occurrence-size distribution function of coronal 
loop widths, characterized by the lower limit of widths $w_{min}$, the peak 
(or most frequent) width $w_p$, the peak occurrence number $n_p$, and a 
power law slope $a$. 
Our data analysis includes automated tracing of curvi-linear features with 
the OCCULT-2 code, automated sampling of the cross-sectional widths of 
coronal loops, and fitting of the theoretical size distribution to the 
observed distribution. With Monte-Carlo simulations and variable pixel 
sizes $\Delta x$ we derive a first diagnostic criterion to discriminate 
whether the loop widths are unresolved $(w_p/\Delta x \approx 2.5\pm0.2)$, 
or fully resolved (if $w_p/\Delta x \gapprox 2.7$). For images with 
resolved loop widths we can apply a second diagnostic criterion that
predicts the lower limit of loop widths as a function of the spatial 
resolution. We find that the loop widths are marginally resolved 
in AIA images, but are fully resolved in Hi-C images, where our model 
predicts a most frequent (peak) value at $w_p \approx$ 550 km, 
in agreement with recent results of Brooks et al. This result agrees 
with the statistics of photospheric granulation sizes and thus supports 
coronal heating mechanisms operating on the macroscopic scale of 
photospheric magneto-convection, rather than nanoflare braiding models 
on unresolved microscopic scales.  
\end{abstract}

\section{		    INTRODUCTION			}

The solar corona is permeated by invisible magnetic field lines,
which can be illuminated by filling with hot coronal plasma, 
also called ``coronal loops'' or ``strands''. They are supposedly field-aligned 
in regions with a low plasma-$\beta$ parameter. The geometry of 
such coronal loops can be characterized by the curvi-linear 3-D coordinates 
$[x(s), y(s), z(s)]$ of a potential or non-potential magnetic field line
(as a function of the loop length coordinate $s$) along the loop axis,
and by the cross-sectional width variation $w(s)$ in transverse direction
to the loop axis. Both properties are controlled by the magnetic field.
Since cross-field transport is inhibited in a low plasma-$\beta$ corona,
plasma can only move along field lines, and thus the spatial organization
of the coronal heating mechanism is to some extent preserved in the
cross-sectional geometry and topology of loop cross-sections. In particular,
we are interested in the size of the thinnest loop strands, which have been
postulated in so-called ``nanoflare heating'' scenarios and may be
detected now with the most recent instruments with the highest spatial 
resolution. Thus we are interested in the spatial organization and statistical 
distributions of loop cross sections, which may provide us clues 
about the geometry of the elusive coronal heating mechanism.  

Previous studies on the width of coronal loop cross-sections have
been focused on the finest widths that could be spatially resolved with a 
particular telescope (i.e., Bray and Loughhead 1985; Loughhead et al.~1985; 
Aschwanden and Nightingale 2005; Winebarger et al.~2013; Peter et al.~2013),
unresolved strands as substructures of loops (Patsourakos and Klimchuk 2007;
Aschwanden and Boerner 2011; Schmelz et al.~2013),
spatial variations of loop widths $w(s)$ along loops (Wang and Sakurai 1998;
Klimchuk 2000; Watko and Klimchuk 2000; 
Lopez-Fuentes and Klimchuk 2010),
temporal loop width variations (Aschwanden and Schrijver 2011),
magnetic modeling of loop width variations (Petrie 2006; Schrijver 2007;
Bellan 2003; Peter and Bingert 2012),
or cross-field transport of energy (Winglee et al.~1988; 
Amendt and Benford 1989; Chae 2002; Ruderman 2003;
Georgoulis and LaBonte 2004; Vasquez and Hollweg 2004;
Voitenko and Goossens 2004, 2005; Galloway et al.~2006; 
Ruderman et al.~2008; Erdelyi and Morton 2009; Pascoe et al.~2009;
Kontar et al.~2011; Bian et al.~2011; Morton and Ruderman 2011;
Arregui and Asensio-Ramos 2014; Arregui et al. 2015;
Kaneko et al.~2015; Ruderman 2015). Summaries 
on the width of coronal loops are given in the textbooks of
Bray et al.~(1991; Chapters 2 and 3) or 
Aschwanden (2004; Chapter 5.4.4).

While theoretical models require the knowledge of the relevant
coronal heating mechanisms, which are still elusive at this time,
there exists essentially no theoretical prediction of the statistical
distribution of loop widths in the solar corona that could be tested 
with current high-resolution data. However, it has been demonstrated
in recent work that the statistical distributions of physical parameters 
in nonlinear energy dissipation processes can be modeled in terms of 
{\sl self-organized criticality (SOC)} models (Bak, Tang, and Wiesenfeld
1987), in geology (e.g., sand piles or earthquakes), as well as 
in a large number of astrophysical phenomena (e.g., solar and stellar 
flares, pulsar glitches, soft X-ray gamma repeaters, blazars, etc.,
for a recent review see Aschwanden et al.~2016). In essence, {\it SOC is a 
critical state of a nonlinear energy dissipation system that is slowly 
and continuously driven towards a critical value of a system-wide 
instability threshold, 
producing scale-free, fractal-diffusive, and intermittent avalanches 
with powerlaw-like size distributions} (Aschwanden 2014). 
If we consider coronal heating events on the most basic level, in the 
sense of an elementary energy dissipation episode in a single loop 
structure, it is not much different in a solar flare or in a nanoflare 
(for scale-free or self-similar processes), and may be produced by the 
same physical mechanisms (of magnetic reconnection, plamsa heating, and 
cooling by thermal conduction and radiative loss), except that flares
have a higher degree of spatial complexity (with multiple loops and strands)
than nanoflares (within a single strand).  
The power law behavior in the temporal and spatial distributions of the
Ohmic heating simulated in a 3D MHD model has been interpreted in terms
of a SOC system specifically for the case of coronal heating 
(Bingert and Peter 2013). Almost every
(nonlinear) instability produces ''avalanching events'' over some 
scale-free range, in contrast to the statistics of linear random events 
that usually form a Gaussian probability distribution with a preferred
typical (spatial) scale. Therefore, the measurement of the statistical
distribution of loop widths should tell us at least whether the
underlying coronal heating mechanism is a linear random process 
(leading to a Gaussian distribution) or a nonlinear dissipative process
(leading to a power law distribution).

A size distribution exhibits a power law function over a restricted
range $[x_1,x_2]$ only (called the ``inertial range'' or 
``scale-free range''), and is bound 
by a lower limit $x_1$ for the smallest observed size, and by an 
upper limit $x_2$ for the largest and rarest observed size, often
close to the system size. The lower limit is affected by the sensitivity 
threshold of the observations and the instrumental spatial resolution. 
The finite spatial resolution modifies the measured width also, in a
predictable manner, as it can be modeled with the point spread function
of the imaging instrument. These effects have never been modeled 
for loop width measurements, which is now possible in large statistical 
samples, thanks to powerful new tools of automated loop recognition 
software that has been developed recently (Aschwanden, De Pontieu, 
and Katrukha 2013a). 

We will use recent high-resolution data from the {\sl Atmospheric 
Imaging Assembly (AIA)} instrument (Lemen et al.~2012) 
onboard the {\sl Solar Dynamics Observatory (SDO)} (Pesnell et al.~2011),
and {\sl Hi-C} data (Kobayashi et al.~2014) 
to test theoretical predictions of loop width distributions.
In Section 2 we start with an analytical description of statistical 
width distributions.
In Section 3 we describe the applied data analysis method, which
includes automated pattern recognition and {\sl Principal
Component Analysis (CPA)} techniques. 
In Section 4 we conduct Monte-Carlo simulations of our automated
loop width detection algorithm, in Section 5 we apply this algorithm
to high-resolution AIA/SDO and Hi-C images, in Section 6
we discuss the results in the light of previous loop width measurements,
and in Section 7 we present the conclusions.

\section{	ANALYTICAL TREATMENT                     }

The physical parameter of interest here is the ``loop width'', which 
we will define in terms of the equivalent width of the radial loop 
cross-sectional flux profile $F(r)$ in Section 3.2 (Eq.~14), 
being equivalent to the full width at half maximum (FWHM).
In order to model size distributions of loop widths in a realistic
manner we have to include threshold effects due to the finite 
sensitivity and spatial resolution of the instruments, which are
characterized in terms of a point spread function. 

\subsection{	The Scale-Free Probability Conjecture 		}

As mentioned in the introduction, the statistics of nonlinear energy 
dissipation events often exhibit a power law-like function in their 
size distribution. A most general testable prediction
that can be made for scale-free nonlinear energy dissipation processes
is the so-called {\sl scale-free probability conjecture} 
(Aschwanden 2012), which universally applies to {\sl self-organized 
criticality (SOC)} systems (Bak et al.~1987), and predicts a power law 
distribution $N(L)$ for length scales $L$,
\begin{equation}	
	N(L) \ dL \propto L^{-D} \ dL \ ,
\end{equation}
where $D$ is the Euclidean dimension of the system. If we use the
widths $w$ of coronal loops as the length scale of the cross-sectional
area over which an ``avalanching'' mechanism dissipates 
energy and heats a particular coronal loop, the
Euclidean dimension of cross-sectional areas is $D=2$, and thus
a power law distribution of $N(w) \propto w^{-2}$ is expected.
The diagram shown in Fig.~1 depicts
3 sets of loops (with widths of $w$=1, 1/2 and 1/4), which are packed 
into 2-D areas with identical sizes ($L=1$), yielding bundles of
$n$=1, 4, and 16 strands, as expected from the size distribution
specified in Eq.~(1). Even when a fraction of the packing area is
used only (e.g., $q=0.25$ as illustrated with dark-grey loop areas
in Fig.~1), the relative probability or size distribution follows
the same statistical power law relationship as specified in Eq.~(1).

\subsection{	Thresholded Power Law Distributions		}

The scale-free range of a size distribution (for instance of widths $x$)
exhibits an ideal power law function for the differential occurrence
frequency distribution, which is characterized with a power law slope $a$, 
\begin{equation}
	 N(x) \ dx \propto x^{-a} \ dx \ .
\end{equation}  
However, there are at least four natural effects that produce a deviation 
from an ideal power law size distribution, which includes: (1) A physical 
threshold of an instability; (2) incomplete sampling of the smallest 
events below a theshold; (3) contamination by an event-unrelated 
background; and (4) truncation effects at the largest events due to a 
finite system size, which all can be modeled with a so-called 
{\sl ``thresholded power law''} distribution function, also called 
Pareto [type II] or Lomax distribution (Lomax 1954),  
\begin{equation}
	N(x)\ dx \propto (x + x_0)^{-a}\ dx \ ,
\end{equation}
which was found to fit flux distributions of solar and stellar flare 
data well, in the scale-free range $x \gapprox x_0$ (Aschwanden 2015). 
The additive constant $x_0$ turns a power law 
function into a constant for small values $x \ll x_0$ at the left side
of a size distribution (Fig.~2a; dashed curve), 
a feature that fits some data, but not all.
Examples for different phenomena are shown in Fig.~8 of Aschwanden (2015),
based on data presented in Clauset et al.~(2009).
The functional shapes of size distributions are shown in Fig.~2, 
which shows the ideal 
power law slope at $x \gg x_0$ (Eq.~2; thin solid line in Fig.~2a), and
the thresholded power law function (Eq.~3; dashed line in Fig.~2a),
where the threshold is set at $x_0/w_{min} = 2.5$, all shown on a log-log 
scale in Fig.~2.

\subsection{	Minimum Loop Widths 		 		}

Since we are going to analyze the contributions of the thinnest 
loop strands to the size distribution of loop widths, we have to
include a number of effects that set a lower limit on the measurement 
of loop widths, such as the finite instrumental spatial resolution and
loop background noise, which we discuss in more detail with a
quantitative example in Section 2.5. At this point we combine these
effects into the variable $w_{min}$, which represents an absolute
lower limit in the theoretical model distribution of loop widths $w$.
The effectively detected loop width (or observed loop width) $w$
can be defined in terms of adding the true loop width $w_{true}$ and the
combined broadening effects expressed in $w_{min}$ in quadrature, 
\begin{equation}
	w = \sqrt{ w_{true}^2 + w_{min}^2 } \ ,
\end{equation}
so that the true width $w_{true}$ can be obtained from the observables
$w$ and $w_{min}$,
\begin{equation}
	w_{true} = \sqrt{ w^2 - w_{min}^2 } \ .
\end{equation}
Substituting the variable $x(w)$ into the thresholded distribution 
function $N(w) dw =N(x[w])$ $|dx/dw| \ dw$ (Eq.~3), where 
we denote $x=w_{true}$ (in Eq.~5) and $w_{min}=x_0$ (in Eq.~(3)), 
and inserting its derivative, $dx/dw=w/\sqrt{w^2-w_{min}^2}$, 
we obtain then the theoretical size distribution $N(w)$ of apparent 
(or modeled observed) loop widths $w$,
\begin{equation}
	N(w) \ dw \propto \left\{
		\begin{array}{ll}
		0	& {\rm for} \ w \le w_{min}	\\
		w \ \left(w^2-w_{min}^2 \right)^{-1/2} 
		\left( \sqrt{(w^2-w_{min}^2)} + w_{min} \right)^{-a} \ dw
			& {\rm for} \ w > w_{min}	
		\end{array}
		\right. \ .
\end{equation}
which is shown in Fig.~2a (dash-dotted line) as a function of the
normalized width $w/w_{min}$. Thus, the spatial resolution causes an upturn 
with an excess above the thresholded power law distribution right near 
$w/w_{min} \gapprox 1$, while no events are detected below the limit 
$w/w_{min} = 1$. However, the sharp peak at $w/w_{min}=1$ may not be observable, 
because the loop width measurements have inherently some uncertainties 
that smooth out such sharp peaks in the observed size distributions,
and show the most frequent detections at a peak value of $w_p/w_{min} 
\approx 2.5$, rather than at the absolute limit $w/w_{min}=1$
(see numerical simulations in Section 4 and point-spread functions of 
EUV imaging instruments, compiled in Table 3 and references therein). 

\subsection{	Power Law with a Smooth Cutoff 		}

In our data analysis,
the width $w$ of a loop cross-section is measured from the equivalent
width of a Gaussian-like loop flux profile, where a local background is
subtracted with a lowpass filter, and the high-frequency data noise is 
filtered out with a highpass filter. This method, as well as many others, 
produce uncertainties in the loop width measurement of order of 
one pixel for the thinnest loop strands, mainly because the true 
background flux profile is unknown and perturbed by data noise. 
As a consequence, the size distribution of loop widths exhibits a 
smooth drop-off towards the minimum value at $w \gapprox w_{min}$. 

In order to provide a realistic
analytical model of a size distribution that can reproduce such an
absolute cutoff value we have to replace the threshold parameter
$w_{\mathrm{min}}$ with a reciprocal function $1/(w-w_{min})$ that has a
singularity at $w=w_{\mathrm{min}}$,
\begin{equation}
	N(w) \ dw = n_0 \left( {w + {w_0^2 \over (w - w_{min}) }} \right)^{-a} 
		\ dw \ ,
\end{equation}
where we introduced an arbitrary constant $w_0$, to be determined.
This type of distribution contains the desired singularity at $w=w_{min}$,
and setting it to zero for smaller values $w < w_{min}$ enforces an
absolute lower cutoff at $w=w_{min}$.

We explore the maximum of this distribution function by setting
the derivative to zero, i.e., $dN/dw|_{w=w_p} =0$, and find that the
peak value $w_p$ of the distribution function (Eq.~7) is related
to the constant $w_0$ by,
\begin{equation}
	w_0 = (w_p - w_{min})	\	,
\end{equation}
while the maximum value $N_p$ of the distribution function at $w_p$
amounts to,
\begin{equation}
	N_p = N(w=w_p) = n_0 (2 w_p - w_{min})^{-a} \ .
\end{equation}
Inserting the constant $w_0$ (Eq.~8) into the distribution function 
$N(w)$ (Eq.~7) yields then,
\begin{equation}
	N(w) \ dw = \left\{
		\begin{array}{ll}
		0	& {\rm for} \ w \le w_{min}	\\
		n_0 \left( {w + {(w_p - w_{min})^2\over (w - w_{min})} } 
		\right)^{-a} \ dw
			& {\rm for} \ w > w_{min}	
		\end{array}
		\right. \ .
\end{equation}
This definition of a size distribution has the following properties,
which are illustrated in Fig.~2:
(1) The distribution has two regimes, a scale-free range 
at large values $(w_p < w < w_{max})$ that follows a power law function,
and a smooth cutoff range at small values $(w_{min} < w < w_p)$;
(2) A peak of the distribution $N_p=n(w=w_p)$ at the value $w=w_p$;
(3) The size distribution is completely defined in the entire range
of $(w_{min} < w < w_{max})$ with an absolute lower cutoff at
$w \approx w_{min}$;
(4) The distribution function can be represented with 
four parameters ($n_0, w_0, w_{min}, a)$, or more conveniently 
expressed in terms of the peak parameters, $(N_p, w_p, w_{min}, a)$,
with $N_p=n_0(2 w_p-w_{min})^{-a}$ and $w_p=w_{min} + w_0$.
Since the parameter $w_{min}$ is a constant for a given 
instrument, only the three parameters $(n_0, a, w_p)$ 
need to be optimized in a fitting procedure for each data set.
A family of such size distributions with various peak values $w_p$
is depicted in Fig.~2b. We will see that this type of power law 
distribution with a smooth cutoff range $[w_{min}, w_p]$ suits 
the observed size distributions much better (Figs.~6h, 10h, 14h)
than the thresholded power law function with the threshold 
constant $x_0 = w_{min}$ (Eq.~3), or the sharply peaked power law 
function with a lower cutoff at the spatial resolution $w_{min}$ 
(Eq.~6).

\subsection{	Loop Width Broadening Effects		}

Let us discuss the physical meaning of the minimum loop width 
$w_{min}$ that we introduced in the previously defined theoretical 
distributions. The major contribution to the minimum detected loop
width is the instrumental point-spread function $w_{psf}$.
Typically, the instrumental point-spread function of EUV or soft X-ray 
telescopes amounts to $w_{psf} \approx 2.0-2.5$ pixels of the CCD detectors, 
mostly dictated by the electronic charge spreading. For instance, the
azimuthally-averaged point-spread function of the Transition Region
and Coronal Explorer (TRACE) has been determined with a blind iterative 
deconvolution technique to be $\approx$2.5 pixels, or $1.25\arcsec$
(Golub et al.~1999). 

In addition, the measured loop widths are affected by data noise 
in the subtracted background counts in every pixel, which is most
severe for faint loops on top of a strongly varying background, especially
since we apply a high-pass filtering technique (rather than a
Gaussian fitting of loop cross-section profiles). From numerous
tests we evaluated that this loop width broadening component has
a standard deviation of $w_{noise} \approx 1$ pixel. A more detailed
discussion of this quantity including Quiet-Sun background counts,
Poissonian photon noise, readout noise, digitization noise,
lossless compression, pedestal dark current, integer subtraction,
and spike residuals is given for the TRACE instrument in 
Aschwanden et al.~(2000c).

Combining these loop width broadening effects by summing in quadrature
we specify an observed loop width $w$ by,
\begin{equation}
	w^2 = w_{true}^2 + w_{psf}^2 + w_{noise}^2 \ ,
\end{equation}
where $w_{true}$ represents the unknown true width for a given loop segment.
A quantitative example is illustrated in Fig.~4e, where a mean FWHM width 
of $w=2.9 \pm 1.1$ pixels is measured from 75 loop cross-sections. 
If we estimate a point-spread function width of $w_{psf}=2.5$ pixels, 
a data noise component of $w_{noise} \approx 1.0$ pixel, we obtain with
Eq.~(11) a true loop width of $w_{true}=
[w^2 - w_{psf}^2 - w_{noise}^2]^{1/2}
\approx [2.9^2 - 2.5^2 - 1.0^2]^{1/2} \approx 1.1$ pixels.
Thus, the expected minimum observed loop width 
for unresolved loop strands (with $w_{true} \ll 1$ pixel) is 
$w_{min} = [w_{psf}^2 + w_{noise}^2]^{1/2} \approx 2.7$ pixels.

In summary, the following symbols are used for loop widths:
$w$ is generally used for the observed (or modeled observed) loop width,
$w_{true}$ is the true (fully resolved) loop width,
$w_{min}$ is the minimum width in the theoretical model distribution,
$w_{sim}$ is the simulated loop width, 
$w_{psf}$ is the point-spread function, and 
$w_{noise}$ is the broadening due to background subtraction data noise.
In the numerical simulations, an instrumental point-spread
function with a width of $w_{psf}=2.5$ pixels is applied.

\section{	DATA ANALYSIS METHOD		}

Our data analysis method consists of two major steps: (1) The automated
loop detection, and (2) a {\sl Principal Component Analysis (PCA)}
applied to generate loop width distributions. 

\subsection{	Automated Loop Detection			}

The automated detection of features with a large range of spatial
scales is not trivial, especially when small and large structures
coexist and blend into each other, or overlap with each other, as it
is the case for a solar corona image with many different loops along
any given line-of-sight.

We make use of an already existing automated loop detection code,
the so-called {\sl Oriented Coronal CUrved Loop Tracing} code OCCULT-2
(Aschwanden et al.~2008c, 2013a; Aschwanden 2010), which is customized 
for automated detection of curvi-linear coronal (or chromospheric) 
loops (or fibrils) with relatively large curvature radii, compared 
with their cross-sectional width. The numerical code, written in the
{\sl Interactive Data Language (IDL)} is publicly available
in the {\sl SolarSoftware (SSW)} library, with a tutorial given at 
{\it http://www.lmsal.com/$\sim$aschwand/software/tracing/tracing$\_$tutorial1.html} .
The OCCULT-2 code can be applied
to any 2-D image with an intensity (or flux) distribution $F(x,y)$,
from which the cartesian coordinates $[x(s), y(s)]$ of a set of
2-D curvi-linear structures as a function of their length 
coordinate $s$ are determined. Applications range from solar EUV  
images of the {\sl Transition Region And Coronal Explorer (TRACE)},
AIA/SDO, H$\alpha$ images of the {\sl Swedish Solar Telescope (SST)},
to microscopic images of microtubule filaments in live cells in
biophysics (Aschwanden et al.~2013a). We provide in the following
a brief description of the OCCULT-2 code. 

The original image is first
filtered with a highpass filter (with a smoothing 2-D box car with a
length of $n_{sm1}$ pixels), and with a lowpass filter (with a smoothing
2-D box car with a length of $n_{sm2}$), which together act as a
highpass filter for curvi-linear structures with a (transverse) length 
scale range of $[n_{sm1}, n_{sm2}]$, where the bipass-filtered image
is defined as,
\begin{equation}
	F_{bipass}(x,y) = smooth{[F(x,y), n_{sm1}]} - smooth{[F(x,y), n_{sm2}]} \ . 
\end{equation}
The highpass filtering of an image is also known as {\sl unsharp masking}.

The numerical algorithm of automated curvi-linear pattern detection  
works as follows. First, the image location $[x_1, y_1]$ of the maximum
flux $F_{max}(x_1, y_1) = max[(F_{bipass}(x,y)]$ is detected, from where the search 
of the first structure (with the highest contrast) starts, by detecting
the directional angle $\alpha([x_1, y_1])$ of the local ridge (Fig.~3), 
\begin{equation}
	\alpha([x_1(s), y_1(s)] = \arctan{\left({dy/ds \over dx/ds}\right)}
	\ .
\end{equation}
In addition, the curvature radius $r_{curv}(x_0, y_0)$
of the local ridge is determined also in order to obtain a second-order
polynomial approximation of the traced loop structure. A geometric
diagram of the parameterized first and second-order elements measured
in a single point of a curvi-linear ridge is shown in Fig.~3. Then the same
direction is extrapolated in second-order as an approximate prediction
for the continuation of the loop direction. The exact direction of the
traced loop segment at the next loop coordinate $s_{i+1}=s_i + \Delta s$ 
is then evaluated by maximizing the cross-correlation coefficient between the 
(generally gaussian-like) subsequent loop cross-sectional profile 
$F[r, \alpha(s_{i+1})]$ compared with the previous profile 
$F[r, \alpha(s_{i})]$. If the direction of the loop tangent is
${\bf s}=[dx/ds, dy/ds]$, the perpendicular direction is defined by
${\bf r}=[-dy/ds, dx/ds]$, along which the cross-sectional loop
profile $F(r)$ is defined. The automated loop detection continues
from $s_i$ to $s_{i+1}=s_i + \Delta x$ in the first loop segment step, 
and then from $s_{i+1}$ to $s_{i+2}$ in the second loop segment step, and 
so forth. The end of the first half loop segment occurs when the
bipass-filtered flux values drop below a selectable threshold
(typically set at one standard deviation of the flux of the 
background variation). 
The code traces then the second half of the loop segment in the same way,
except in reverse direction, starting from the initial maximum value
at $[x_1,y_1]$. Once the end of the second half loop segment
is reached, both the first and second half segments are joined together
in the same direction, which constitutes the coordinates 
$[x(s_i), y(s_i)], i=0,...,n_p$ with $n_p$ points of the full loop
segment. The image area that covers the coordinates of the first loop,
with a margin of $\pm n_{sm1}$ pixels in $x$ and $y$-direction,
is then erased in the bipass-filtered image, so that an already traced 
loop segment is not used multiple times in the tracing of a new 
structure. This procedure of detecting the first
loop is then continued at the location of the next image maximum 
$[x_2, y_2]$ and repeated for the structure with the 
second-highest contrast, and so forth, until the zero floor or a
preset threshold value in the bipass-filtered image is reached. 
In the end the code produces 
a list of $[x_j(s_i), y_j(s_i)]$ loop coordinates for $j=1,...,n_{j}$
loop segments, each one containing $i=1,....,n_i$ coordinate points. 

There are a few control parameters that can be set in the
OCCULT-2 code, such as 
the maximum number of traced structures $n_{struc}$ per image or 
wavelength, the highpass filter $n_{sm1}$, the lowpass filter $n_{sm2}$,
the minimum accepted loop length $l_{min}$, the minimum allowed
curvature radius $r_{min}$, the field line step $\Delta s$ along
the (projected) loop coordinate, the flux threshold $q_{thresh1}$
(in units of the median value of positive fluxes in the original
image), the filter flux threshold $q_{thresh2}$ (in units
of the median value in the positive bipass-filtered fluxes),
the maximum allowed gap $n_{gap}$ with zero or negative fluxes along
a traced ridge. In our analysis we choose the following default control
parameters: $n_{sm1}=1,...,64$ pixels, $n_{sm2}=n_{sm1}+2$ pixels,
$l_{min}=2$ pixels, $r_{min}=100$ pixels, $\Delta s=1$ pixel, 
$q_{thresh,1}=0$, $q_{thresh,2}$=2, and $n_{gap}=0$.
We will present some parametric studies in Section 4 in order to 
investigate systematic trends of the algorithm.

\subsection{	Principal Component Analysis			}

The OCCULT-2 code applies
a bipass-filter to an image $F(x,y)$, consisting of a highpass
filter with a selectable smoothing boxcar length $n_{sm1}$, and
a lowpass filter with a boxcar length $n_{sm2}$. Such a bipass 
filter selects only structures within the width range of $[n_{sm1}, 
n_{sm2}]$ in the automated pattern recognition algorithm.
The narrowest possible width for a bipass filter is 
$n_{sm2}=n_{sm1}+2$, which yields a highest sensitivity to
detect structures with a width of $w \approx (n_{sm1}+n_{sm2})/2$. 
For instance, if we set a lowpass filter of $n_{sm1}=2$, the
highpass filter is $n_{nsm2}=n_{sm1}+2=4$, and the selected
loop widths are $w \approx (n_{sm1}+n_{sm2})/2 = (2+4)/2=3 \pm 1$.
An example of an automatically traced loop with extraction
of the loop width profiles in perpendicular direction to the
loop axis is shown in Fig.~4. The actual loop width $w$ at a
given loop location $s_i$ along a loop is measured from the
equivalent width $w$ of the loop cross-sectional flux profile
$F(r)$, 
\begin{equation}
	w = {\int F(r) dr \over max[F(r)] } \ ,
\end{equation}
where the coordinate $r$ is defined in perpendicular
direction to the local loop axis $s$, and the value $F_{ij}=F(s_i, r_j)$
at a particular location $(s_i, r_j)$ is obtained from bilinear
extrapolation in the 2-D image $F(x,y)$. In the example shown
in Fig.~4e, the mean loop width is found to be $w=2.9\pm1.1$ pixels,
where the uncertainty mostly results from the point-spread function,
the noisy background, and multiple overlapping loop strands.
The automated measurement of loop widths is performed in steps
of $\Delta s = 1$ pixels along each loop segment, which typically
yields 10-100 times more width measurements than loop detections.

For the detection of structures with cross-sectional widths
covering a range of two decades, we can select a series of 
independent filter width components that are increasing by 
powers of two,
\begin{equation} 
	n_{sm1,k} = 2^k \ , 
	\qquad k = 0, 1, ... 6 \ ,
\end{equation}
which essentially corresponds to the orthogonal basis functions 
in Fourier 
transforms or in {\sl Principal Component Analysis (PCA)} techniques 
(e.g., Jackson 2003). PCA techniques have been applied to solar
physics data sets in various ways (e.g., Lawrence et al.~2004;
Casini et al.~2005; Cadavid et al.~2008; Zharkova et al.~2012).
Hence, we can run the automated loop detection for every image
multiple times with independent (orthogonal) bipass filters $n_{sm1,k}$,
in order to obtain a completely sampled range of width measurements.
The smallest smoothing boxcar width ($n_{sm1,0}=1$ pixel) approximately 
corresponds to the instrumental resolution, while an upper limit of
$n_{sm1,6}=2^6=64$ pixels covers about two orders of magnitude in
the range of width measurements, with $w_{max} = 2^6 = 64$ pixels.

The choice of filters by powers of 2 yields a uniform distribution
of widths $w_i, i=0,...,6$ on a logarithmic scale, $log{(w)}$.
The OCCULT-2 code yields then a number of detected loop structures
$N_i, i=0,...,6$ for each selected width $w_i$. This yields directly
a frequency occurrence distribution (or size distribution) $N(w)$ 
of loop structures, which is generally displayed on a 
$log(N)-log(S)$ (Number versus Size) plot. 
In other words, the option of width filters in the OCCULT-2 code 
can be directly used for the measurement of size distributions $N(w)$.
Of course, structures that are detected on relatively large scales
of $w \gg 1$ Mm, are likely to consist of more complex structures (e.g., 
arcades of loops), rather than traditional field-aligned monolithic loops.
The size distributions can have different functional shapes, each one
being a characteristic of the underlying physical process. For instance,
random measurements of linear processes yield a Gaussian normal
distribution, while nonlinear dissipative processes yield a 
power law distribution. By measuring the size distribution, we obtain
thus a diagnostic of the underlying (linear or nonlinear)
physical generation mechanism.

\section{	MONTE-CARLO SIMULATIONS			}

In order to test and validate the numerical code to measure the size
distributions $N(w)$ of loop widths $w$ within a specified range,
we present a basic Monte-Carlo simulation (Section 4.1), and
investigate the effect of finite spatial resolution on the 
inferred loop width size distribution $N(w)$ (Section 4.2).  

\subsection{	Basic Monte-Carlo Simulation		}

We generate a thresholded power law size distribution of loop widths,
with a threshold at $w_0$ (according to Eq.~3), where the symbol
$w_{sim}$ stands here for the simulated fully resolved loop widths,
\begin{equation}
	N(w_{sim}) \ dw_{sim} \propto (w_{sim} + w_0)^{-a} \ dw_{sim} 
	\qquad {\rm for}\ w_{sim} \ge w_{min} \ ,
\end{equation}
with a power law index of $a=2$, bound by the range 
$[w_{min},w_{max}]$. Such a distribution can be produced by using 
the transformation (e.g., Section 7.1 in Aschwanden 2011;
Eq.~15 in Aschwanden 2015),
\begin{equation}
	w_{sim,i} = \left[(1 - \rho_i)(w_{min}+w_0)^{(1-a)}
	+ \rho_i (w_{max} + w_0)^{(1-a)} \right]^{(1/(1-a)} - w_0 \ ,
\end{equation}
where $\rho_i$ are random numbers uniformly distributed
in the interval $[0 < \rho_i < 1]$. Such a set of values $w_{sim,i}$
forms then a size distribution $N(w_{sim})$ of widths in form of a 
power law function. The peak value of the cross-sectional 
loop profiles $F(r)$ are chosen according to the
emission measure definition in an optically thin plasma,
\begin{equation}
	F_i \propto \int n_e^2\ dz \propto \langle n_e^2 \rangle w_{sim,i} 
	\propto w_{sim,i} \ ,
\end{equation}
where the mean electron density $\langle n_e \rangle$
is assumed to be independent of the
loop width, which means that the flux increases proportionally
to the column depth $w = \int dz$, being equal to the loop width $w_{sim,i}$
for circular loop cross-sections. Note that the simulated loop widths
$w_{sim,i}$ defined in Eqs.~(16-18) are fully resolved, because each
simulated loop strand is made up of numerical sub-strands with a width
of 1 pixel.

A spatial 2-D image $F(x,y)$ is then composed by superposing the
brightness distributions of semi-circular loops (as shown in Fig.~5), 
each one consisting
of a space-filling bundle of loop strands with a total 
Gaussian cross-section of width $w_{sim,i}$ (Eq.~17) and flux $F_i$ (Eq.~18). 
The loop positions are randomly distributed in 2-D space 
with a fixed height of the curvature center (of the semi-circular loops)
in photospheric height. 
In this set up we simulate a total of $n=128$ loop bundles,
where the strands have a cutoff or minimal width of $w_0=1$ pixel,
projected into a 2-D image with a size of $1000 \times 1000$ pixels. 
In addition, the simulated images are convolved with a 2D Gaussian 
kernel that has a FWHM of $w_{psf}=2.5$ pixels in order to mimic 
a realistic instrumental point-spread function.
The peak count rate is $10^5$ photons per pixel, and a background
with random noise of 10\% is added. 
An example of such a simulated image is shown in Fig.~5a, 
along with a bipass-filtered rendering in Fig.~5b. The
added random noise is visible in the filtered image in Fig.~5b.

We perform now a validation test of the capability to retrieve the
underlying power law distribution function of loop widths as follows. 
We apply seven bipass filters with $n_{sm1} = 2^i = 
1, 2, 4, 8, 16, 32, 64$ and $n_{sm2}=n_{sm1}+2$ to produce seven
bipass-filtered images with different filter widths.
This corresponds to the principal component analysis (PCA) 
described in Section 3.2. We run the automated loop detection code
OCCULT-2 on each of the 7 images, using a threshold of $q_2=2.0$
standard deviations (in the bipass-filtered image), we require
a minimum length of $l_{min}=2$ pixels for the selected loop length
segments, and show the detected loop tracings in Fig.~6 (red curves), 
which appear to be fairly complete in all filters, as judged by 
visual detection. Comparing the detected structures with the
theoretically simulated image (Fig.~5a) on a pixel-by-pixel basis,
we find that 80\% of the simulated image pixels are retrieved
with the correct values, using the automated OCCULT-2 code. 
Then we fit the histograms
of loop widths, which includes 152,564 loop widths measurements in
1254 loop segments detected in 7 different filter images.  
For the power law slopes of the resulting width distributions
we find the values of $a=2.82\pm0.07$ for the differential frequency 
distribution (Fig.~6h), and a similar value from the cumulative 
frequency distribution (Fig.~6i). 

The specific value of the power law slope $a$ cannot easily be 
retrieved with our method, since most
loops show up in multiple width filters, and thus are counted
multiple times with different widths. We tested input values
of the power law slope in the range of $a=1.5-4.0$ 
in the simulations, but obtained invariably values of 
$a \approx 2.7 \pm 0.2$ after auto-detection with 
our multi-scale sampling method. 
A method for the power law slope retrieval can in principle be
designed by eliminating multiple countings of loops, but this
requires a unique definition of the mathematical function of
loop cross-section profiles, which is an ambiguous task and 
is not necessary for the measurement of the size distribution 
of the smallest loop strands here.

Inspecting the peak of the size distribution of widths $N(w_{sim})$, 
we find a peak width of $w_p/w_{min}=3.0$ (Figs.~6h, 6i).
The reduced chi-square value of $\chi = 1.9$ (Fig.~6h) is
satisfactory for the fit of the differential size distribution,
given the simple analytical model used here (Eq.~10). 

Thus this test demonstrates that we can (1) measure
a complete size distribution of loop widths within a range
of about two orders of magnitude above the instrumental
resolution limit, (2) retrieve the functional shape of a power law 
function (although not the exact value of the power law slope), 
and (3) detect the rollover of a smooth cutoff in the range of 
$1 \le w_{sim}/w_{min} \lapprox 3$ pixels near the resolution limit 
$w_{min}$. Comparing simulated loop images (Fig.~6) with
observed images (Figs.~13-14), however, we notice that
the numerical simulations contain idealized loops only, without
taking temperature inhomogeneities into account, such as ``moss
structures'' at the footpoints of hot soft X-ray emitting loops
(e.g., Berger et al.~1999). The present study therefore does not
discriminate between ``classical coronal loops'' and ``loop-like
transition region phenomena'', nor does it disentangle the
3D topology of nested loop structures. Nevertheless, the sampling
is complete in the 2D image plane (thanks to the automated
loop detection algorithm), and thus no ad hoc assumptions 
of loop selection criteria are applied, which makes it more 
suitable to compare with theoretical models of statistical 
distributions, rather than a hand-picked sample of selected 
(loop) structures.

\subsection{	Detecting the Cutoff of the Smallest Loop Strands	}

The detection of a lower cutoff for the widths of the smallest 
loop strands depends crucially on the spatial resolution or
pixel size of the image.
In a Monte-Carlo simulation we can vary the spatial resolution
and study the effect of the spatial resolution limit on the
detection of the finest loop strands. In the basic example
shown in Fig.~6 we have chosen a minimum loop strand width 
of $w_{min}=1$ image pixel, and sampled the size distribution
in the range between $w_{min}=1$ and $w_{max}=10^2$ pixels 
(Figs.~6h and 6i). We note that the resulting differential
occurrence frequency distribution shown in Fig.~6h exhibits
a peak at a width ratio of $w_p/w_{min} \approx 3.0$. It turns out
that this ratio of the peak width to the pixel size
is a persistent feature in our simulations of power law
distributions $N(w_{sim})$ of loop widths, for samples with unresolved
structures down to the pixel size $\Delta x$. 

In order to investigate the universality of this ratio of the
peak value $w_p$ of a size distribution $N(w_{sim})$ to the
pixel size $\Delta x$ we perform 20 numerical simulations
of identical 2-D images (the same as shown in Fig.~5a),
but with 20 different pixel sizes, from $\Delta x=0.1 \arcsec$
to $\Delta x=2.0$ \arcsec, while the scaling of the standard simulation
shown in Fig.~6 corresponds to an AIA pixel with $\Delta x = 0.6\arcsec$.
For conversion into units of Mm we use the scaling $1\arcsec$ (arcsec)=0.725 Mm.
We overlay the histograms and power law fits of the 20 size 
distributions with different pixel sizes in Fig.~7a, which shows a 
systematic shift of the peak with pixel size. However, when we normalize
the size distributions $N(w_{sim})$ to the pixel size $\Delta x$ on the x-axis,
and to the peak value $N_p=N(w_{sim}=w_p$) on the y-axis, we see that the peaks
line up at a peak value of $w_p \approx 3.0 \Delta x$ (Fig.~7b), and thus
the normalized size distribution of loop widths exhibits a
universal shape. However, this is only true for distributions
that do not resolve the finest loop strands (red curves in Fig.~7), 
when the smallest loop width is smaller than the pixel size, 
while the distributions with 
pixel sizes smaller than the finest loop widths exhibit
a larger ratio $w_p/\Delta x \gapprox 3.0$ of the peak width to the pixel 
size, because there is a relative scarcity of detected loops
at small widths $w_{sim}$. Thus we can use the ratio $w_p/\Delta x$
as a diagnostic for discriminating which size distributions 
resolve, or do not resolve, the smallest loop strands.    

In order to quantify this new diagnostic criterion we perform $4 \times 20$
Monte-Carlo simulations, for four different cutoffs $w_{min}$
of the loop width distributions, $w_{min}=0.3\arcsec$, 
$0.6\arcsec$, $1.2\arcsec$, and $1.8\arcsec$, and for 20
different spatial resolutions $\Delta x=0.1\arcsec,
0.2\arcsec, ..., 2.0\arcsec$. For each of the 80 simulated
images we repeat the automated loop detection, produce
the size distributions of loop widths, and measure the ratio
of the peak width to the pixel size, $w_p/\Delta x$. The
results of these ratios as a function of the pixel size
are plotted in Fig.~8 (black diamonds).
From the four plots in Fig.~8 we see that each function
$w_p/\Delta x$ exhibits two regimes, a fully resolved
regime at $w < w_{min}$ on the left, and an unresolved regime
at $w > w_{min}$ on the right side. We can model this dependence of
the peak width $q_w=w_p/\Delta x$ on the spatial resolution $\Delta x$
with the function $q_w(\Delta x)$,
\begin{equation}
	q_w = \sqrt{ \left({w_{min} \over \Delta x} \right)^2 
		+ \left( w_{psf} \right)^2 } \ ,
\end{equation}
for which best fits are shown in Fig.~(8) (red curves), yielding
approximate values for the point-spread function ($w_{psf} \approx 2.5$ 
pixels) and the minimum width $w_{min}$ for each of the 4 simulations.
A more accurate value for the minimum width $w_{min}$ is obtained
by interpolating the fitted function at the critical value
$q_w=3.0$ pixels, which is the separator between the fully resolved
and unresolved regimes. For the 4 cases simulated in Fig.~8 we used
minimum loop widths of $w_{min}=0.3\arcsec$, $0.6\arcsec$, $1.2\arcsec$, 
and $1.8\arcsec$, while the predicted values inferred from the critial
value $q_w=3.0$ are $w_{min}^{pred}=0.35\arcsec$, $0.51\arcsec$, 
$0.72\arcsec$, and $2.07\arcsec$, and thus agree with the values 
used in the numerical simulation within $\lapprox 30\%$.  

In summary, our numerical simulations demonstrate that 
rebinning an image with different pixel sizes and sampling of the resulting
size distributions of loop widths is capable to predict the
minimum width $w_{min}$ at the cutoff of the finest loop strands
contained in the image, using the diagnostic of the peak width 
ratio to the pixel size, $w_p/\Delta x$.  The applied
range of rebinned pixel sizes should not extend below the
instrumental pixel size of the image, because there is no 
information on finer scales in the image. If the finest structures
are all unresolved at full resolution, the peak width
ratio has a constant value of $w_p/\Delta x \lapprox 3.0$.
If the peak width ratio is larger, this indicates that the finest 
loop strands are resolved and the critical limit $\Delta x$ or 
minimum loop width $w_{min}$ can be determined from the inversion of
$q_w (\Delta x)$ (Eq.~19). This diagnostic 
gives us a reliable tool to determine the size $w_{min}$ of the
smallest loop strands in the corona, in the case of resolved structures, 
or an upper limit in case of unresolved structures.
If this cutoff is detected on a macroscopic scale, we can conclude 
that we hit ``the rock bottom of the smallest loop strands''.  

\section{	DATA ANALYSIS OF OBSERVATIONS			}

In the following we analyze EUV images from AIA/SDO (Section 5.1),
and from the Hi-C rocket flight (Section 5.2). 
Since we are interested in the finest detectable loop strands in the 
solar corona, these two data sets are selected because of their 
highest spatial resolution that is available from simultaneous
EUV observations.

\subsection{	AIA/SDO					}

We analyze AIA images from 2011 February 14, 00:00 UT,
in the six coronal wavelengths 94, 131, 171, 
193, 211, and 335 \ang . The AIA images have a full image size
of 4096$\times$4096 pixels, with a pixel size of 0.6\arcsec,
corresponding to a spatial resolution of $w_{res}=0.6\arcsec
\times 2.5$ pixels $\times$ 0.725 Mm $\approx$ 1.1 Mm 
on the solar surface).  The time cadence
of AIA is 12 s. For the automated pattern recognition we cut out
a subimage with a field-of-view of 0.3 solar radii (or
487$\times$487 pixels), centered on
the active region NOAA 11158, which has a heliographic position
of S20 and E27 to W38 during the observed 6 days. One example
is shown for 2011 Feb 14 in Figs.~9 and 10. The same active region is
subject of over 40 published studies (for a list of references
see Section 5.1.1. in Aschwanden, Sun, and Liu 2014). 
Descriptions of the SDO spacecraft and the AIA instrument can 
be found in Pesnell et al.~(2011) and Lemen et al.~(2012).  

In each AIA image (at 6 wavelengths) we run the 
automated loop recognition code OCCULT-2 separately, sampling 
the number of loops ($n_{loop}$) and detected loop widths ($n_{wid}$),
and performed power law fits for both the differential and
cumulative size distributions. The results of the
total number of detected loops $n_{loop}$, the power law
slopes $a_{diff}$ and $a_{cum}$, and the goodness-of-fit 
$\chi^2$-values are listed in Table 1.

When we compare the results from AIA (Fig.~10) with those of
our numerical simulation (Fig.~6), we find similar values for
the power law slope $a \approx 3.1$, and the peak width 
$w_p \approx 2.9$ pixels, which confirms that our design
of Monte-Carlo simulations closely reproduces the functional 
shape of the observed coronal loop width distributions. A new insight 
of this study is how the spatial resolution affects the 
occurrence frequency distribution $N(w)$ in the range from $w=1$ pixel
to the peak of the size distribution of $w_p \lapprox 2.9$ pixels,
which can be diagnosed from the normalized peak width value
$w_p/\Delta x$.

For the power law slopes of the loop width distributions,
averaged over all AIA wavelengths, we found 
mean values of $a_{diff}=2.7\pm0.3$ and $a_{cum}=2.3\pm0.4$ (Table 1),
which corroborates the consistency between the differential and 
cumulative power law fitting method. The variation of the power
law slope among different wavelengths is of order 10\% (Table 1).
The measured power law slopes $a$ can be understood as the
fractal dimension of the geometric volume of an active region,
because the simulated loop bundles are not space-filling,
although the strands inside a loop bundle are space-filling.
If an active region would be
solidly filled with loops, we would expect a fractal dimension
that corresponds to an Euclidean dimension of $a=D=3$.
On the other side, if all detected loops are located in a
2-D layer with a fixed width in the third dimension, for
instance caused by gravitational stratification, the expectation
would be a Euclidean dimension of $a=D=2$. The fact that we find 
a power law index with a mean value $a \approx 2.7\pm0.3$
indicates that the spatial distribution of measured loop
cross-sections is fractal. In comparison, a fractal dimension of 
$D_3=2.0\pm0.5$ has been inferred from 3-D modeling of 20 solar 
flare events (Aschwanden and Aschwanden 2008a,b).

\subsection{	Hi-C 			 		}

We turn now to coronal images with the highest spatial resolution
ever recorded, which were obtained during a rocket flight by
the {\sl High-resolution Coronal Imager (Hi-C)} on 2012 July 11.
Instrumental descriptions are available from Kobayashi et al.~(2014)
and Cirtain et al.~(2013). We use an image recorded at 18:54:16 UT, 
in the wavelength band of 193 \ang , which is dominated by the
Fe XII line originating around $T \approx 1.5$ MK. The image
has a size of $3880 \times 4096$ pixels, with a pixel size of
$0.1\arcsec$ (corresponding to 73 km/pixel), covering a
field-of-view of 300 Mm ($\approx 0.37 R_{\odot}$) squared, 
and was sampled with an exposure time of 2 s. The same image 
was analyzed previously by Peter et al.~(2013).

We present the analyzed image in Fig.~11a, which contains a
sunspot (in the upper half of the image), moss regions
(in the upper left quadrant), as well as coronal loops 
in the periphery of an active region (bottom right quadrant),
which are analyzed in Peter et al.~(2013).

The structures hidden in the image can be enhanced by bipass
filtering (Eq.~12), as revealed in Fig.~12a with a bipass
filter of $n_{sm1}=16$ and $n_{sm2}=18$, which is most
sensitive to structures with a width of $w=17 \times 0.1\arcsec
\times 725$ km = 1200 km, or in Fig.~12b with a bipass
filter of $n_{sm1}=32$ and $n_{sm2}=34$, which is most
sensitive to structures with a width of $w=33 \times 0.1\arcsec
\times 725$ km = 2400 km. However, filtering at the
highest resolution, as shown in Fig.~11a with a bipass filter
of $n_{sm1}=1$ and $n_{sm2}=3$, being most sensitive to 
structures with a width of $w=2 \times 0.1\arcsec \times 725$ km 
= 145 km, reveals no significant structures above the
noise level anywhere (Fig.~11b), which is a surprising result that we
will scrutinize in the following analysis in more detail.

In Fig.~13 we show the automated loop detection applied to
the $3880 \times 4096$ pixel full resolution HiC image, using
the 7 bipass filters from $n_{sm1}=2$ to $n_{sm2}=128$.
Almost no significant structure is seen at the highest
resolution (Fig.~13a and 13b), while a maximum of 
$N_{det}=661$ structures is detected at the filter
$n_{sm1}=32$ (Fig.~13e). The resulting size distribution of
widths shows a peak width of $w_p/\Delta x =7.1$ pixels (Fig.~13h),
which according to our numerical simulations is indicative
of fully resolved width structures at the smallest scale
of $w_{min}=1$ pixel or $0.1\arcsec$, because we would expect a
peak width of $w_p/\Delta x \approx 2.5$ for unresolved structures
at full resolution. 

\subsection{	Hi-C and AIA Comparisons 		}

A comparison of the Hi-C image with the simultaneous and 
cospatial AIA image at the identical wavelength of 193 \ang \
has been studied in Peter et al.~(2013), from which we use
the same images, which have been coaligned and corrected
for a rotation of $1.9^\circ$. The only difference between
these two images is the pixel size, with $0.1\arcsec$ for
Hi-C, and $0.6\arcsec$ for AIA, being a factor of 5.8 
different. AIA images have been taken 3 s before and 6 s
after the Hi-C image. The results of our inter-comparison
are tabulated in Table 2.

We perform an identical analysis of the AIA image (Fig.~14)
as we did for the Hi-C image (Fig.~13). In the AIA image,
significant structures are already visible at the full
resolution (Fig~14a), which are comparable with those
seen in the Hi-C image in the $n_{sm1}=17$ filter (Fig.~13d).
Inspecting the size distribution of loop widths obtained
in the AIA image, we find also a different peak width
of $w_p/\Delta x = 3.1$ (Fig.~14h), which indicates according
to our numerical simulations that the finest structures
are barely resolved in the AIA image at its full resolution
($0.6\arcsec$), while they appear to be fully resolved
in the Hi-C image at its full resolution ($0.1\arcsec$).

In Fig.~15 we show a juxtaposition of the size distributions of
the AIA image with an original pixel size of $0.6\arcsec$ (Fig.~15a), 
the AIA image degraded to the Hi-C resolution of 
$0.1\arcsec$ (Fig.~15d), the Hi-C image with the original
pixel size of $0.1\arcsec$ (Fig.~15c), and the
Hi-C image re-scaled to the AIA resolution of $0.6\arcsec$
(Fig.~15b). We find that the size distributions of widths 
are almost identical for equal pixel sizes, such as AIA
and the rescaled Hi-C image with $0.6\arcsec$ pixels 
(Figs.~15a and 15b), or the Hi-C and the rescaled AIA image 
with $0.1\arcsec$ pixels (Figs.~15c and 15d). This means
that there is essentially the same information stored for
equal pixel sizes if all structures are resolved at this
pixel size, which however would not be the case for unresolved
structures. Further we find that rescaling of the same
instrument does not preserve the peak width $w_p$, or
the normalized peak width $w_p/\Delta x$, as it can be
seen in the comparison of AIA (Fig.~15a) and the rescaled
AIA (Fig.~15d), or in the comparison of Hi-C (Fig.~15c)
and the rescaled Hi-C image (Fig.~15b). The results of this
inter-comparison are summarized in Table 2. This is all
consistent with our method that the peak width ratio
$w_p/\Delta x$ can be used in the rescaling of the same image
as a diagnostic for discriminating whether the finest
structures in an image are resolved or not.

In the results of the numerical simulations presented
in Fig.~8, we demonstrated that the peak width
ratio has a constant value of $w_p/\Delta x \approx 2.5$ 
for unresolved structures when the image is rescaled to
different pixel sizes $\Delta x$, while the ratio increases 
to higher values at the highest resolution (of one pixel
size) for images with fully resolved structures.  
For this experiment we rebin the Hi-C image from a pixel
size of $\Delta x=0.1\arcsec$ to $\Delta x=2.0\arcsec$ 
in 20 equidistant steps of $\Delta x_i = 0.1 \times i, 
i=1,...,20$. Thus the rescaled image varies in the number
of pixels from $3880 \times 4096$ pixels down to 
$194 \times 205$, but covers the same field-of-view and
thus the same structures. We repeat for each of the 20
rescaled Hi-C images the automated pattern recognition
algorithm and the sampling of width histograms, from
which we extract the peak width ratio $w_p/\Delta x$
and plot it as a function of the pixel size $\Delta x$
(Fig.~16a). We find that the ratio is near-constant 
for large pixel sizes, while it increases to smaller
pixel sizes. The separation point between fully resolved
and unresolved loop widths can be read from the critical
value $w_p/\Delta x=3.0$ on the y-axis, which yields
a value of $\Delta x=0.77\arcsec$ on the horizontal
axis, corresponding to $w_p \approx 550$ km. Thus we 
conclude that the structures detected with Hi-C have 
the most frequent width value of $w_p\approx 550$ km, 
while thinner loop strands becoming rapidly sparser 
below this limit.

We repeat the same experiment with the AIA image 
for full-resolution (down to 1 pixel) and
show the results in Fig.~16b. If we employ the same
critial value of $w_p/\Delta x=3.0$ on the y-axis, 
we can read of a value of $\Delta x = 0.58$ \arcsec,
which corresponds to $\Delta x = 420$ km, which 
agrees with the value from Hi-C $\Delta x = 550$ km
within $\approx 25\%$.
Thus, the AIA data indicate that all structures are
mostly unresolved, but become marginally resolved
at the full resolution of 1 AIA pixel,
i.e., $\Delta x \approx 0.6\arcsec$. 
Only with the additional information of Hi-C we can
conclude that most structures are resolved below the
AIA resolution.

The ratio of the detected (or observed) loop width $w$ to the true 
(or simulated) loop width $w_{true}$ follows from Eq.~(4), 
i.e., $w = \sqrt{w_{true}^2 + w_{min}^2}$,
as visualized in Fig.~17. From Fig.~17 we can read off
the ratio of the peak widths $w_p$ to the minimum width $w_{min}$,
which is found to be $w_p/w_{min}=6.4$ for AIA, and 
$w_p/w_{min}=19.4$ for Hi-C, while we would expect a ratio of
$w_p/w_{min} \approx 2.5$ for instruments that do not resolve
the loop strands. Thus the diagram shown in Fig.~17 
tells us that Hi-C fully resolves the bulk of loop strands,
while AIA resolves most of them too, but only by a small margin
compared with a non-resolving instrument.

\section{	DISCUSSION				}

\subsection{	Comparison with Previous Loop Width Measurements 	}

We provide a comprehensive compilation of 52 studies 
that dealt with coronal
loop width measurements in Table 3, with a graphical visualization 
in Fig.~18. Summaries of loop width measurements can also be
found in Bray et al. (1991; Chapters 2 and 3), Aschwanden (1995),
and Aschwanden (2004; Chapter 5.4.4). The graphical representation
in Fig.~18 divides the loop width measurements in 4 wavelength
regimes by color (optical + H$\alpha$ + Ly$\alpha$, EUV, 
soft X-rays, and radio), and ranks the width ranges by the
smallest detected width in ascending order. There are several
trends visible in this overview. Minimum loop widths have been
measured from $w \approx 20$ Mm down to $w_{min} \approx 0.1$ Mm.
The finest loop widths have been detected preferentially 
in EUV, while the
loop widths measured in soft X-rays and optical wavelengths
tend to be significantly larger, and are found to be largest in
radio wavelengths. This can be explained because coronal loops 
in EUV and soft X-rays are produced by optically thin 
bremsstrahlung, which yields a better contrast than the partially 
optically thick free-free and gyroresonance emission in radio 
wavelengths.

In Table 3 we also list the pixel size $w_{pix}$ and the spatial
resolution $w_{res}$ of the instruments. In Fig.~18 we can clearly
see at one glance that the smallest loop widths are almost always
measured a significant factor above the instrumental pixel sizes,
which is partially explained by the point-spread function that
typically amounts to 2.5 pixels in current EUV imagers (TRACE,
AIA/SDO, STEREO, IRIS, Hi-C). Interestingly, the same ratio
of $w_p/\Delta x \approx 2.5$ is also found in our present work,
which corresponds to the optimum detection efficiency of our
automated loop tracing algorithm (using bipass filters)
in the case of the Monte-Carlo simulations, besides the effects
of the point-spread function in the analyzed AIA and Hi-C data. 
Since every loop width measurement
is subject to noise in the background, which causes some scatter
in the order of a pixel size, for instance $w \approx 2.9\pm1.1$  
(pixels) in the example shown in Fig.~4. The lowest width measurements
can be as low as $w \gapprox 1$ pixel using the equivalent-width
method (Eq.~14). Some measurements shown in Fig.~18 exhibit even
smaller widths than one pixel, because the authors attempted to
determine the true width after deconvolution of the point-spread
function, i.e., $w_{true} \approx \sqrt{w^2 - w_{res}^2}$ (Eq.~4 
and Fig.~17).
 
Virtually all studies that report loop width measurements 
are based on a single loop or a small number of visually selected
(hand-picked) loops, which may contain the smallest features
visible in an image, but they are not representative samples
that would allow us to derive the entire width distribution 
of all coronal loops seen in an image. Our work thus represents
a unique method that completely samples the entire size distribution
function $N(w) dw$ over a range of about two orders of magnitude,
characterized by the peak value ($w_p$), a smooth cutoff range
$[w_{min}, w_p]$ down to the minimum width value 
$(w_{min})$, and by a power law function
in the upper part $[w_p, w_{max}]$ up to the maximum size 
$(w_{max})$. Consequently, the range of loop widths measured
in the present study ($w \approx 0.1-10$ Mm) entails all previous
EUV measurements of loop widths, as it can be seen at one
glance in Fig.~18. 

However, we may ask whether the distribution
of loop sizes continues at the low end if a future instrument
faciliates a higher spatial resolution. Based on our peak width
diagnostic criterion, this can only be the case if we measure
a value of $w_p \approx 2.5\pm0.2$ for the peak in the size
distribution with the current highest-resolution instruments.
The Hi-C measurements with 0.1\arcsec resolution, based on 
$n_{wid}=138,965$ individual loop width measurements (Fig.~13) 
indicate a peak width ratio of $w_p/\Delta x=7.1$ (Fig.~13h), 
which clearly suggests that most detected structures are fully
resolved at $\Delta x=0.1\arcsec$ resolution, while the most
common loop width (which is given by the peak $w=w_p$ in the
size distribution), amounts to $\Delta x=0.77\arcsec$ (or
$\Delta x \approx 550$ km; Fig.~16a). This does not exclude the
possible measurement of smaller loops, but smaller loops 
are expected to be less common than loops with a mean width of
$w_p \approx 550$ km, according to our criterion. Smaller loops
have indeed been measured from recent Hi-C studies,
e.g., $w=200-1500$ km (Peter et al.~2013),
$w=117-667$ km (Brooks et al.~(2013),
$w=150-310$ km (Morton and McLaughlin 2013),
or $w=120-150$ km (Brooks et al.~(2016).
Our prediction is that loops at $w \approx 550$ km are the most
frequently occurring loops and form the peak of the size
distribution, if sufficient spatial resolution is available,
requiring $\Delta x \le w_p/2.5 \approx 120$ km (or $\approx 0.16\arcsec$).
This implies that a statistical analysis of IRIS data 
(De Pontieu et al.~2014) should be
able to reproduce the peak at $w_p \approx 550$ km (or 0.77\arcsec).
Most interestingly, Brooks et al.~(2013) show an occurrence
distribution of 91 hand-picked loops observed with Hi-C with a minimum
Gaussian width of $\sigma_w= 90$ km (FWHM=212 km) and an average 
Gaussian width of $\sigma_w=272$ km (FWHM=640 km),
which is fully consistent with our result of a peak 
width (FWHM) of $w_p\approx 550$ km based on three orders of magnitude
larger statistics. Winebarger et al.~(2014) conducted a 
statistical analysis of the noise characteristics of a Hi-C image
and concluded that 70\% of the pixels in each Hi-C image do not
show evidence for significant substructures, which confirms the
scarcity of thinner loop strands than observed so far (see also
Fig.~11b, where no structures except data noise is visible).
Moreover, Peter et al.~(2013) demonstrate that some of the loops 
seen in Hi-C are essentially resolved with AIA and appear to have 
a smooth Gaussian-like cross-section.

\subsection{	Consequences for Coronal Heating Models 	}

There are two schools of coronal heating models: 
(1) The macroscopic view considers monolithic coronal loops that 
have heating cross-sections commensurable with the size of 
magneto-convection  cells ($w \approx 300-1000$ km), and thus 
are mostly resolved with current EUV imagers (TRACE, AIA, 
IRIS, Hi-C); and (2) the microscopic view that postulates unresolved 
nanoflare strands that cannot be resolved with current instrumentation. 
Our continuously developing technology has faciliated
tremendous improvements in astronomical high-resolution observations,
so that the available spatial resolution improved by a factor of about 300
over the last 35 years. In particular, the measurement of
coronal loop diameters improved from $w \approx 30$ Mm (Davis and
Krieger 1982) down to $w \approx 0.1$ Mm with the current Hi-C observations
(Peter et al.~2013; Brooks et al.~2013, 2016). Our new finding
of a preferred spatial scale of $w_p \approx 550$ km for coronal loops 
calls for a physical explanation of this particular value. 
This peak width value $w_p$ separates the two regimes of a
smooth cutoff range ($w < w_p$) and of a scale-free power law
distribution at larger values ($w > w_p$). 

One obvious physical scale is the granulation structure of the
photosphere, discovered by Sir William Herschel in 1801.
The spatial size of the time-varying granulation is measured
by a combination of high-pass and low-pass filtering in the
spatial frequency domain with subsequent thresholding, which
segments granular and intergranular lanes. Alternative
procedures are ``lane-finding'' schemes that are based on local
gradient detection (Spruit et al.~1990). With the latter method,
the smallest granules, near the observational limit, were found
to have a most frequent size of 290 km. The size distribution 
peaks at $w_p=290$ km, then shows a slow decline
in number up to a diameter near 1500 km, and then a rapid
decline towards the largest granule sizes (Brandt 2001). 
Modern measurements with the {\sl New Solar Telescope at Big Bear},
show a peak in the occurrence distribution of granule sizes 
shifted further down to $w \approx 150$ km (Abramenko et al.~2012).

If we adopt the magneto-convection of photospheric granular cells
as the primary driver of local magnetic reconnection processes
(in the chromosphere, transition region, and corona), for instance
see numerical MHD simulations by Gudiksen et al.~(2005) 
or Bingert and Peter (2013),
we would expect a congruence between the magnetic reconnection
area and the cross-sectional loop heating area (Aschwanden et al.~2007a,b). 
The diffusion region of a local magnetic reconnection process essentially
defines the spatial scale of a local heating event, and therefore
the cross-sectional area of a coronal loop. 
Our new result of a preferred cross-sectional scale of $w_p \approx
550$ km, which appears to be perfectly comparable with the most
frequent granular scale, adds an additional constraint to the geometry 
of the heating process, besides ten other arguments that have been 
discussed earlier in the context of the coronal heating paradox 
(Aschwanden et al.~2007a,b).

The second school of thought deals with unresolved ``microscopic''
structures. The ``nanoflare heating'' scenario (Klimchuk 2006) 
builds on Parker's (1988) conjecture
that tangential misalignments of the magnetic field between adjacent
flux tubes sporadically reconnect, driven by the braiding random
motion of coronal loop footpoints in the photosphere. Since
magneto-convection of the photosphere plays an important role
as driver for magnetic braiding, which applies to both schools,
the major difference between
the monolithic and the nanoflare heating scenario is mostly the
assumed spatial scale of the heating region, which contrasts 
the macroscopic view versus the microscopic view. There is a general
consensus that broad loops are likely to consist of bundles of
finer strands, as alluded to in Fig.~1, but the crucial question 
remains what the finest scale of a composite loop is. In the
scale-free regime of a width distribution, a power-law like
distribution function is generally observed that extends from
$w_{max}$ down
to the peak value $w_p$, where it breaks down by some unknown
cutoff mechanisms. If nanoflares on finer scales exist, we would
expect that the power law function extends to lower values than
the observed cutoff, while the apparent peak width occurs at
$w_p/\Delta x \approx 2.5$ due to the finite resolution. However, 
our crucial observation of $w_p/\Delta x \gg 2.5$ in Hi-C data
contradicts this assumption that most frequent nanoflare strands 
exist at smaller spatial scales than $w < 550$ km. This implies a physical
limit, similar to the observed lower limit of granule sizes.
Therefore we conclude that the Hi-C data are not consistent with
a nanoflare scenario with most frequent loop strands at $w_p < 550$ km. 
In addition, current nanoflare scenarios have difficulties to explain the
isothermality of macroscopic loops (Aschwanden and Nightingale 2005;
Aschwanden 2008), but see recent modifications in terms of
``nanoflare storm'' scenarios (Klimchuk 2015).

The best efforts to model Parker's nanoflare scenario do not show 
a clear structuring across the (large-scale) magnetic field. In the 
turbulent (reduced) MHD model of Rappazzo et al.~(2008) current 
sheets (along the large-scale magnetic field) form throughout 
the computational domain.  No obvious structuring of the current 
sheets into bundles are found, where the heating is concentrated 
and which (in a more realistic model) could give rise to a coronal 
loop. So there is the need for an external process that would lead 
to a cross-field scale producing loops of finite width, and the 
photospheric magneto-convection is a good candidate for this.

\subsection{	Consequences for Coronal Density Measurements 	}

Besides the consequences of loop width statistics on coronal heating
models, there are numerous implications for coronal electron density
measurements also. The emission measure of optically thin plasma, observed in
EUV and soft X-rays, depends on the column depth, which is set equal to loop
diameters for coronal loops with a circular cross-section. Thus, the density
in coronal loops can only be determined if their diameter or width is known,
such as SXT/Yohkoh measurements of soft X ray-bright flare loops 
(Aschwanden and Benz 1997).
Therefore, measurements of coronal widths are essential for all theoretical
models that require an electron density, such as
calculations of the thermal energy, coronal filling factors, 
gyro-synchrotron emissivity, Alfv\'en speed, or coronal seismology. 
Such consequences will be discussed elsewhere.

\section{	CONCLUSIONS				}

In this study we explore the distribution of coronal loop widths with 
the AIA/SDO and the Hi-C instrument, which provide the highest spatial
resolution available among current solar EUV imagers. Particular 
emphasis is given to the simultaneous data gathered during the 
Hi-C rocket flight on 2012 July 11.
The scientific motivation for this study is the investigation of limits
on the finest loop strands, which represent the ``rosetta stone'' 
of coronal heating models in discriminating microscopic (nanoflare
strands) versus macroscopic (monolithic) loops. For this purpose we
calculate analytical functions for the expected size distribution
of loop widths, conduct numerical Monte-Carlo simulations to develop
a diagnostic tool to discriminate whether the finest detected loop
strands are spatially resolved or not, and analyze AIA and Hi-C data to 
apply this diagnostic tool to find fundamental limits on the 
smallest loop strands. We briefly summarize the results and 
conclusions in the following.

\begin{enumerate}
\item{\underbar{Automated loop detection:} We apply the OCCULT-2
code for the automated detection of curvi-linear structures in
EUV images from AIA and Hi-C. This code is able to measure
$\approx 10^5$ loop widths in a Hi-C image. While previous loop
widths have been measured only in small numbers and produced
non-representative samples, this code allows us to obtain complete 
size distributions of loop widths, extending over two orders
of magnitude, which was not possible with previous manual methods.}

\item{\underbar{Principal Component Analysis:} We applied 
bipass filters (consisting of a low-pass and high-pass filter),
with a series of independent filter width components that increase
by a factor of two over two orders of magnitude, which corresponds
to orthogonal basis functions in principal component analysis
methods. This range of spatial scales is suitable to reconstruct
power law-like size distributions of loop widths.}

\item{\underbar{The size distribution of loop widths:} While no
coronal loop model exists that predicts the size distribution of
loop widths, we make use of the self-organized criticality approach,
which predicts a power law function with a slope that depends on
the (spatial) fractal dimension of a nonlinear energy dissipation system.
We generalize the ideal power law function by including threshold
effects and corrections due to finite instrumental spatial resolution.
We introduce a smooth (lower) cutoff, which leads to a new analytical
formulation of the size distribution function over the entire range
$[w_{min}, w_{max}]$, characterized by 4 free parameters: the
minimum width $w_{min}$, the peak width $w_p$, the peak number $n_0$, 
and the power law slope $a$.}

\item{\underbar{The peak width diagnostic:} From Monte-Carlo simulations
of EUV images with numerous loops that follow a prescribed power law
distribution with a lower cutoff and random fluxes we explore how
the peak width (or most frequently occurring width) $w_p$ depends
on the lower cutoff in the width distribution of loop strands, using 
different pixel sizes $\Delta x$ (to mimic different spatial resolutions). 
We find that loop distributions that have a peak at $w_p/\Delta x 
\approx 2.5\pm0.2$ contain unresolved strands, while larger
width ratios indicate resolved structures at a pixel size $\Delta x$. 
In the case of resolved structures, we find a relationship between
the minimum strand width $w_{min}$ and the critical resolution $\Delta x$
that separates the two regimes of resolved and unresolved loop widths,
i.e., $w_p/\Delta x = 3.0$, which can be used to diagnose the lower cutoff of a 
completely sampled loop width distribution.}

\item{\underbar{Marginally resolved loop widths with AIA:} 
The AIA 193 \ang\ image 
rescaled to pixel sizes ranging from $0.6\arcsec$ to $20.0\arcsec$ yields
size distributions of loop cross-section widths that have a peak at an
invariant ratio of $w_p/\Delta x = 2.59 \pm 0.16$, while a slightly higher
value is obtained near the full resolution, which indicates according to our 
numerical simulations that the structures are marginally resolved
at the full resolution of $\Delta x = 0.6\arcsec$, and that the most
frequent value of the loop size distribution has a value of
$w_{min} \gapprox 420$ km.}

\item{\underbar{Resolved loop widths with Hi-C:} The Hi-C 193 \ang\
image that was observed simultaneously with the AIA 193 \ang\ image,
sampled at full resolution of $\Delta x = 0.1\arcsec$, yields a size 
distribution of loop widths peaking at $w_p/\Delta x = 7.1$, which
indicates according to our numerical simulations that all structures
are over-resolved at the full resolution. When we re-scale the Hi-C
image in the range of $\Delta x=0.1\arcsec-20.0\arcsec$, we find a
most frequent loop width of $\Delta x \approx 550$ km.}

\item{\underbar{Comparison with previous loop measurements:} We identified
52 studies that contain loop width measurements, observed in optical,
H$\alpha$, Ly$\alpha$, soft X-rays (SXR), extreme ultra-violet (EUV), 
and radio wavelengths, which report values over a range of $w=0.1-30$ Mm. 
All previous measurements were carried out
on single loops or a small number of loops, while this study samples
for the first time loop cross-sections over a scale range of two 
orders of magnitude in each image. The finest loop widths have been
detected with IRIS and Hi-C have a lower limit of 
$w_{min} \gapprox 100$ km and a peak value of the most frequently
observed (resolved) loop width of $w_p \approx 550$ km. The closest
confirmation is the Hi-C study of Brooks et al.~(2013), which shows
91 loops with a low cutoff of $w \gapprox 200$ km and a peak at
$w_p \approx 640$ km.}

\item{\underbar{Consequences for coronal heating models:} The coronal
loop widths inferred here, with a peak value of $w_p \approx 550$ km
and a minimum of $w_{min} \approx 100$ km agrees with the size of
the most frequently occurring granule sizes in the photosphere.
We conclude that this geometric congruence supports coronal heating
models that are driven by photospheric magneto-convection and
operate most frequently on a macroscopic scale of $w \approx 550$ km, 
rather than nanoflare heating models that involve unresolved 
loop strands beyond the spatial resolution of current instrumentation.}

\end{enumerate}

This is the first study that finds an absolute lower limit on the
width of coronal loops, based on automated measurements with large
statistics, from which the size distribution of loop widths can be 
sampled and diagnostic criteria can be derived that are capable to
distinguish whether the finest loops observed in a high-resolution
image are resolved or not, for a given instrument. Since a size
distribution contains much more information than a single case
measurement, this statistical approach with automated loop detection 
and size distribution modeling offers more powerful tools to test
coronal heating models than previous single point measurements,
which all are not representative samples of the entirity of all coronal 
loops. If the inferred absolute lower limit on loop widths holds up
in complementary future studies, we may conclude that we are hitting
the rock bottom of loop width sizes. A major consequence may be
the irrelevance of (hypothetical) unresolved nanoflare structures.

\bigskip
\acknowledgements
We acknowledge useful comments from an anonymous referee and 
discussions with Harry Warren, Jim Klimchuk, Karel Schrijver,
and Amy Winebarger. Part of the work was supported by the
NASA contracts NNG04EA00C of the SDO/AIA instrument and
NNG09FA40C of the IRIS mission.


\clearpage

\begin{deluxetable}{rrrrrrr}
\tablecaption{Measurement of the power law slope 
of loop width distribution functions obtained in 
different wavelengths of the AIA image of 2011 Feb 14, 00:00 UT: 
The number of detected loops $n_{loop}$, measured widths $n_{wid}$,
the power law slope of the differential frequency distribution $a_{diff}$,
and of the cumulative frequency distribution $a_{cum}$,
and the goodness-of-fit values $\chi_{diff}$ and $\chi_{cum}$.}
\tablewidth{0pt}
\tablehead{
\colhead{Wave}&
\colhead{Number}&
\colhead{Number}&
\colhead{Power law}&
\colhead{Power law}&
\colhead{Goodness}&
\colhead{Goodness}\\
\colhead{length}&
\colhead{of loops}&
\colhead{of widths}&
\colhead{slope}&
\colhead{slope}&
\colhead{of fit}&
\colhead{of fit}\\
\colhead{}&
\colhead{$n_{loop}$}&
\colhead{$n_{wid}$}&
\colhead{$a_{diff}$}&
\colhead{$a_{cum}$}&
\colhead{$\chi_{diff}$}&
\colhead{$\chi_{cum}$}\\
\colhead{[\ang ]}&
\colhead{}&
\colhead{}&
\colhead{}&
\colhead{}&
\colhead{}&
\colhead{}}
\startdata
  94 &  80 &  2653 & 2.5$\pm$0.3 & 1.8$\pm$0.2  & 0.8 & 1.0 \\
 131 & 199 &  5796 & 2.7$\pm$0.2 & 2.6$\pm$0.2  & 1.0 & 2.2 \\
 171 & 463 & 15017 & 2.7$\pm$0.1 & 2.8$\pm$0.1  & 1.1 & 1.9 \\
 193 & 401 & 11589 & 3.1$\pm$0.2 & 2.7$\pm$0.1  & 1.1 & 1.0 \\
 211 & 310 &  9232 & 2.7$\pm$0.2 & 2.2$\pm$0.1  & 0.8 & 2.7 \\
 335 & 125 &  3871 & 2.4$\pm$0.2 & 1.9$\pm$0.2  & 0.8 & 1.1 \\
     &     &       &             &              &     &     \\
Mean &     &       & 2.7$\pm$0.3 & 2.3$\pm$0.4  &     &     \\
\enddata
\end{deluxetable}

\begin{deluxetable}{lrrrrrrrr}
\tablecaption{Measurement of the power law slope 
of loop width distribution functions obtained from 
simultaneous HiC and AIA 193 A images scaled to each others pixel size:
the number of detected loops $n_{loop}$, measured widths $n_{wid}$,
the normalized peak width $w_p/\Delta x$,
the power law slope of the differential frequency distribution $a_{diff}$,
and of the cumulative frequency distribution $a_{cum}$,
and the goodness-of-fit values $\chi_{diff}$ and $\chi_{cum}$.}
\tablewidth{0pt}
\tablehead{
\colhead{Instrument}&
\colhead{Pixel}&
\colhead{Number}&
\colhead{Number}&
\colhead{Peak}&
\colhead{Power law}&
\colhead{Power law}&
\colhead{Goodness}&
\colhead{Goodness}\\
\colhead{}&
\colhead{size}&
\colhead{of loops}&
\colhead{of widths}&
\colhead{width}&
\colhead{slope}&
\colhead{slope}&
\colhead{of fit}&
\colhead{of fit}\\
\colhead{}&
\colhead{}&
\colhead{$n_{loop}$}&
\colhead{$n_{wid}$}&
\colhead{$w_p$}&
\colhead{$a_{diff}$}&
\colhead{$a_{cum}$}&
\colhead{$\chi_{diff}$}&
\colhead{$\chi_{cum}$}\\
\colhead{}&
\colhead{(arcsec)}&
\colhead{}&
\colhead{}&
\colhead{}&
\colhead{}&
\colhead{}&
\colhead{}&
\colhead{}}
\startdata
AIA     & 0.6 &  784 &  20,927 & 3.1 & 2.6$\pm$0.1 & 2.2$\pm$0.1 & 1.7 & 1.3 \\
HIC \& AIA pixels & 0.6 &  770 &  20,629 & 2.9 & 2.6$\pm$0.1 & 2.3$\pm$0.1 & 1.7 & 1.7 \\
HIC     & 0.1 & 2459 & 138,965 & 7.1 & 1.4$\pm$0.1 & 2.0$\pm$0.1 & 1.5 & 4.8 \\
AIA \& HIC pixels & 0.1 & 3205 & 188,926 & 8.3 & 2.7$\pm$0.1 & 2.5$\pm$0.1 & ... & ... \\
        &     &     &         &                    &             &     &     \\
Mean    &     &     &         &      & 2.3$\pm$0.6 & 2.3$\pm$0.2 &     &     \\
\enddata
\end{deluxetable}
\clearpage

\begin{deluxetable}{lrrrlll}
\tablecaption{Compilation of coronal loop width measurements 
in chronological order.} 
\tablewidth{0pt}
\tablehead{
\colhead{Object}&
\colhead{Pixel}&
\colhead{Spatial}&
\colhead{Loop}&
\colhead{Wavelength}&
\colhead{Instrument}&
\colhead{Ref.\tablenotemark{1}}\\
\colhead{}&
\colhead{size}&
\colhead{resolution}&
\colhead{width}&
\colhead{}&
\colhead{}&
\colhead{}\\
\colhead{}&
\colhead{$w_{pix}$ [Mm]}&
\colhead{$w_{res}$ [Mm]}&
\colhead{$w_{loop}$ [Mm]}&
\colhead{$\lambda$ [\ang]}&
\colhead{}&
\colhead{}}
\startdata
green corona loops	& ...	& ...	& $3-8$	     & 5303       & Dunn/SacPeak   & 1       \\
large green corona loops& ...	& ...	& $8-12$     & 5303	  & Dunn/SacPeak   & 2       \\
green corona loops 	& ...	& 3.6	& $11-22$    & 5303	  & Pic-du-Midi    & 3       \\
red corona structures   & ...	& 3.6	& $8-20$     & 6374       & Pic-du-Midi    & 3       \\
cool loops 		& ...	& 3.6	& $2-5$	     & 1032 O VI  & Skylab	   & 4       \\
cool loops 		& ...	& 3.6	& $<2$	     & 465 Ne VII & Skylab	   & 4       \\
hot loops 	        & ...	& 3.6	& $3-12$     & 1032 Mg X  & Skylab	   & 4       \\
hot loops 		& ...	& 3.6	& $3-12$     & 465 Si XII & Skylab	   & 4       \\
loop prominence 	& ...	& 1.5	& $6.5-7.2$  & 160-630    & Skylab         & 5       \\
active region loop	& ...	& 1.5	& $7-11$     & Fe XVI	  & Skylab 	   & 6       \\
radio loop structure 	& ...	& 5.3	& $22$       & 11 cm      & NRAO           & 7       \\
radio loop structure 	& ...	& 1.7	& $9$        & 3.7 cm     & NRAO           & 7       \\
active region loops	& ...	& 3.1	& $20$       & 20 cm      & VLA            & 8       \\
active region core loops& 2.2	& ...	& $<2.2-5$   & SXR        & ASE rocket     & 9       \\
compact volume loops    & 2.2	& ...	& $5-15$     & SXR        & ASE rocket     & 9       \\
outward extending loops & 2.2	& ...	& $10-30$    & SXR        & ASE rocket     & 9       \\
hot coronal loops 	& ...	& 1.8	& $8-14$     & 170-630    & Skylab         & 10      \\
coronal loops		& ...	& 3.1	& $16-20$    & 20 cm      & VLA            & 11      \\
active region loops	& ...	& 0.7	& $0.93-2.1$ & H$\alpha$  & CSIRO	   & 12,13   \\
very thin loops (chrom) & ...	& 0.7	& $0.44-0.58$& H$\alpha$  & CSIRO 	   & 14      \\
very thin loops (photo) & ...	& 0.7	& $<0.13-0.32$&H$\alpha$  & CSIRO          & 14      \\
coronal loops 		& 0.6	& 0.7	& $2-3.5$    & Ly$\alpha$ & LMSAL rocket   & 15      \\
X-ray loops		& ...   & 1.4   & $6-13$     & 8-65       & ASE rocket     & 16      \\
soft X-ray loops  	& 3.5 	& ...	& $7-9$	     & SXR	  & SXT/Yohkoh     & 17      \\
soft X-ray flare loops	& ...   & 1.8   & $4.2-18.4$ & SXR        & SXT/Yohkoh     & 18      \\
active region loop	& 1.9   & ...   & $5.8$      & 171, 195   & EIT/SOHO       & 19      \\
active region loops 	& 1.9	& ...	& $7.1\pm0.8$& 171        & EIT/SOHO	   & 20      \\
active region loops     & 1.9	& ... 	& $7.8\pm0.8$& 195        & EIT/SOHO	   & 21      \\
active region loops 	& 1.9	& ...	& $7.9\pm1.4$& 284        & EIT/SOHO	   & 21      \\
nanoflare loops 	& 0.35	& 0.82	& $1.8-6.1$  & 171, 195   & TRACE 	   & 22      \\
postflare arcade loops  & 0.35  & 0.82  & $0.9$      & 171, 195   & TRACE          & 23      \\
soft X-ray loop   	& 3.5 	& ...	& $21$	     & SXR	  & SXT/Yohkoh     & 24      \\
oscillating loops	& 0.35  & 0.82  & $8.7\pm2.8$& 171, 195   & TRACE          & 25      \\
nanoflare loops 	& 0.35	& 0.82	& $3-10$     & 171, 195   & TRACE 	   & 26      \\
nanoflare loops 	& ... 	& 1.8 	& $7-50$     & SXR        & SXT/YOHKOH 	   & 26      \\
oscillating loops 	& 0.35	& 0.82	& $5.1\pm3.9$& 171, 195   & TRACE 	   & 27      \\
cooling loops 		& 0.35	& 0.82	& $1.8-12.0$ & 171, 195   & TRACE 	   & 28      \\
heated loop		& 0.35  & 0.82  & $3-6$      & 171, 195   & TRACE          & 29      \\
elementary loops	& 0.35  & 0.82  & $1.4\pm0.3$& 171, 195, 284 & TRACE       & 30,31   \\
cool $(<0.1$ MK) loops  & 0.09  & 0.22  & $0.7-2.2$  & 1216 & VAULT    & 32\\ 
cooling loops 		& 0.35	& 0.82	& $<0.35-8.0$  & 171, 195 & TRACE 	   & 33      \\
3D-reconstructed loops	& 1.15  & 2.30  & $2.6\pm0.1$& 171,195,284& EIT/STEREO     & 34,35   \\
coronal loops           & 0.35  & 0.82  & $0.7-2.2$  & 171,195    & TRACE          & 36      \\
3D-reconstructed loops	& 1.15  & 2.30  & $<1.0\pm1.0$& 171,195,284& EIT/STEREO    & 37      \\
coronal loops           & 0.44	& 0.98	& $4.7-9.7$  & 94-335     & AIA/SDO 	   & 38      \\
oscillating loop        & 0.44	& 0.98	& $4.9\pm0.6$& 94-335     & AIA/SDO 	   & 39      \\
cooling loops 	        & 0.35	& 0.82	& $1.29-1.50$& 171, 195   & TRACE 	   & 40      \\
coronal loops     & 0.73  & ...   & $0.7-1.9$  & 195        & EIS/Hinode& 41\\
coronal rain      & 0.06  & 0.14  & $0.31$     & 6563       & CRISP    & 42\\
auto-detected loops	& 0.44	& 0.98	& $3-8$      & 94-335     & AIA/SDO 	   & 43      \\
coronal loops		& 0.07  & 0.22  & $0.2-1.5$  & 193        & Hi-C           & 44      \\
coronal loops		& 0.07  & 0.22  & $0.117-0.667$& 193      & Hi-C           & 45      \\
inter-moss loops	& 0.07  & 0.22  & $0.675-0.803$& 193      & Hi-C           & 46      \\
oscillating loops	& 0.07  & 0.22  & $0.15-0.31$& 193        & Hi-C           & 47      \\
filament threads        & 0.07  & 0.22  & $0.58\pm0.07$& 193      & Hi-C           & 48      \\
finely structured corona& 0.07  & 0.22  & $>0.22$    & 193        & Hi-C           & 49      \\
coronal rain      & 0.06  & 0.14  & $0.15-0.28$& 6563       & CRISP  & 50\\
coronal rain      & 0.08  & 0.15  & $0.32-0.57$& 3968       & SOT/Hinode&50\\
coronal rain      & 0.12  & 0.30  & $0.44-0.72$& 2796       & IRIS     & 50\\
coronal rain      & 0.12  & 0.30  & $0.40-0.62$& 1330       & IRIS     & 50\\
coronal rain      & 0.12  & 0.30  & $0.46-0.70$& 1400       & IRIS     & 50\\
coronal rain      & 0.44  & 0.98  & $1.02-1.24$& 304        & AIA/SDO  & 50\\
coronal rain      & 0.44  & 0.98  & $0.51-1.52$& 304        & AIA/SDO  & 50\\
coronal rain      & 0.44  & 0.98  & $1.19-2.36$& 171        & AIA/SDO  & 50\\
coronal rain      & 0.44  & 0.98  & $1.17-2.73$& 193        & AIA/SDO  & 50\\
fine structure loops	& 0.12  & 0.30  & $0.12-0.15$& 1400       & IRIS           & 51      \\
penumbral jets 		& 0.07  & 0.22  & $<0.6$     & 193        & Hi-C           & 52      \\
\enddata
\tablenotetext{1}{References: 
(1)  Kleczek (1963); 
(2)  Dunn (1971);
(3)  Picat et al. (1973);
(4)  Foukal (1975);
(5)  Cheng (1980);
(6)  Cheng, Smith, and Tandberg-Hanssen (1980);
(7)  Kundu, Schmahl, and Gerassimenko (1980);
(8)  Kundu and Velusamy (1980);
(9)  Davis and Krieger (1982);
(10) Dere (1982);
(11) Lang, Willson, and Rayrole (1982);
(12) Loughhead and Bray (1984);
(13) Loughhead, Bray, and Wang (1985);
(14) Bray and Loughhead (1985);
(15) Tsiropoula et al. (1986), Transition Region Camera (TRC);
(16) Webb et al. (1987);
(17) Klimchuk et al. (1992);
(18) Aschwanden and Benz (1997);  
(19) Aschwanden et al. (1998);  
(20) Aschwanden et al. (1999);
(21) Aschwanden et al. (2000a); 
(22) Aschwanden et al. (2000b); 
(23) Aschwanden and Alexander (2001); 
(24) Aschwanden (2002);
(25) Aschwanden et al. (2002);
(26) Aschwanden and Parnell (2002);
(27) Aschwanden et al. (2003a);
(28) Aschwanden et al. (2003b);
(29) Petrie et al.~ (2003);
(30) Aschwanden and Nightingale (2005);
(31) Aschwanden, Nightingale, and Boerner (2007);
(32) Patsourakos et al. (2007);
(33) Aschwanden and Terradas (2008);
(34) Aschwanden et al.~(2008a);
(35) Aschwanden et al.~(2008b);
(36) Lopez Fuentes et al. (2008);
(37) Aschwanden and W\"ulser ~(2011);
(38) Aschwanden and Boerner (2011);
(39) Aschwanden and Schrijver (2011);
(40) Mulu-Moore et al. (2011);
(41) Brooks et al. (2012);
(42) Antolin and Rouppe van der Voort (2012);
(43) Aschwanden et al. (2013b);
(44) Peter et al.~(2013);
(45) Brooks et al.~(2013);
(46) Winebarger et al.~(2013);
(47) Morton and McLaughlin (2013);
(48) Alexander et al. (2013);
(49) Winebarger et al.~(2014);
(50) Antolin et al.~(2015);
(51) Brooks et al. (2016);
(52) Tiwari et al. (2016). - 
For summaries see Bray et al. (1991), Aschwanden (1995), 
and Aschwanden (2004; Chapter 5.4.4).}
\end{deluxetable}
\clearpage


\begin{figure}
\plotone{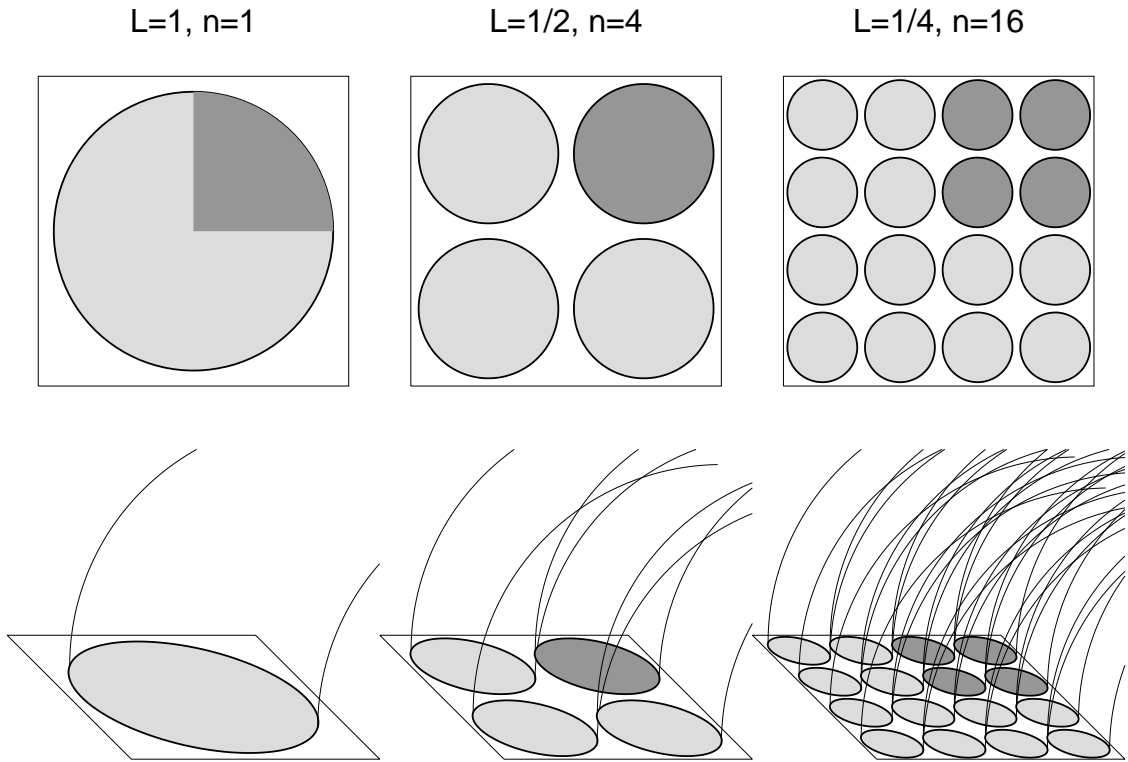}
\caption{2-D view (top panels) and 3-D view of an array of loop
cross-sections that illustrate the reciprocal relationship between
the loop width $w$ and the number $n(w)$ of loops that can be packed
into a given area with a size of $L=1$. For the half size $L=1/2$ 
(middle panels),
a total of $n=4$ loops can be fitted in. For the quarter size $L=1/4$
(right panels), a total of $n=16$ loops can be fitted in. Even when
the active area of loop heating (dark grey areas) has a smaller
fraction than unity ($q=0.25$ here), the relative probability
of active loops with a given width $w$ is invariant.}
\end{figure}
\clearpage

\begin{figure}
\plotone{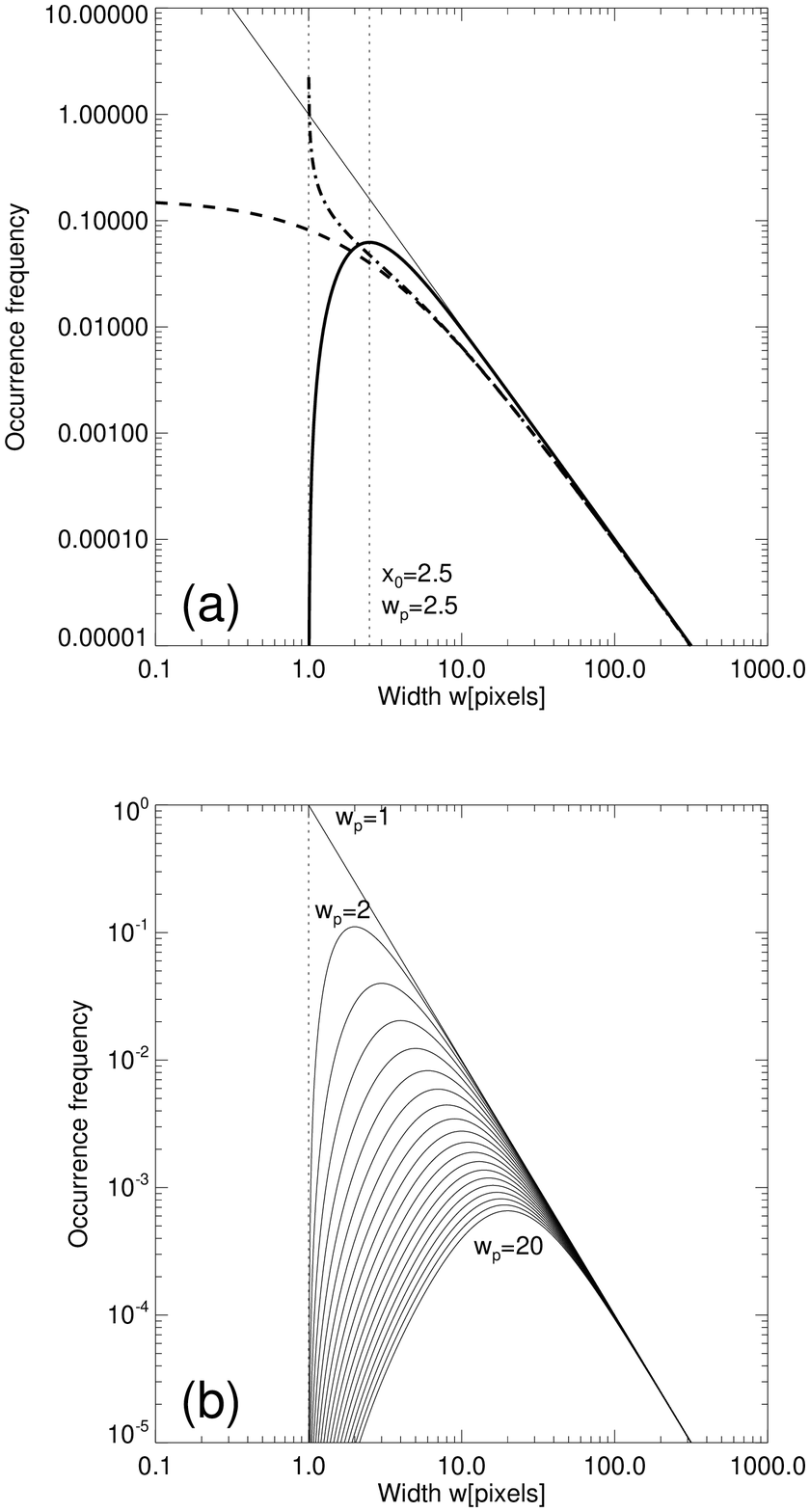}
\caption{(a) Analytical function of an ideal power law (thin solid line),
a thresholded power law with a threshold $x_0/w_{min}=2.5$ 
(dashed curve), a thresholded power law with a finite spatial resolution 
$w_{\mathrm{min}}$ (dash-dotted curve), and a power law function with
a smooth cutoff peaking at $w_p/w_{min}=2.5$ according to Eq.~11 
(thick solid line). The minimum value at $w_{min}$ and the sampling 
threshold at $x_0/w_{min}=2.5$ are indicated with vertical dotted lines.
(b) A family of power law functions with smooth cutoffs and
peaks at $w_p/w_{min}=1, 2, ..., 20$ (Eq.~11).}
\end{figure}

\begin{figure}
\plotone{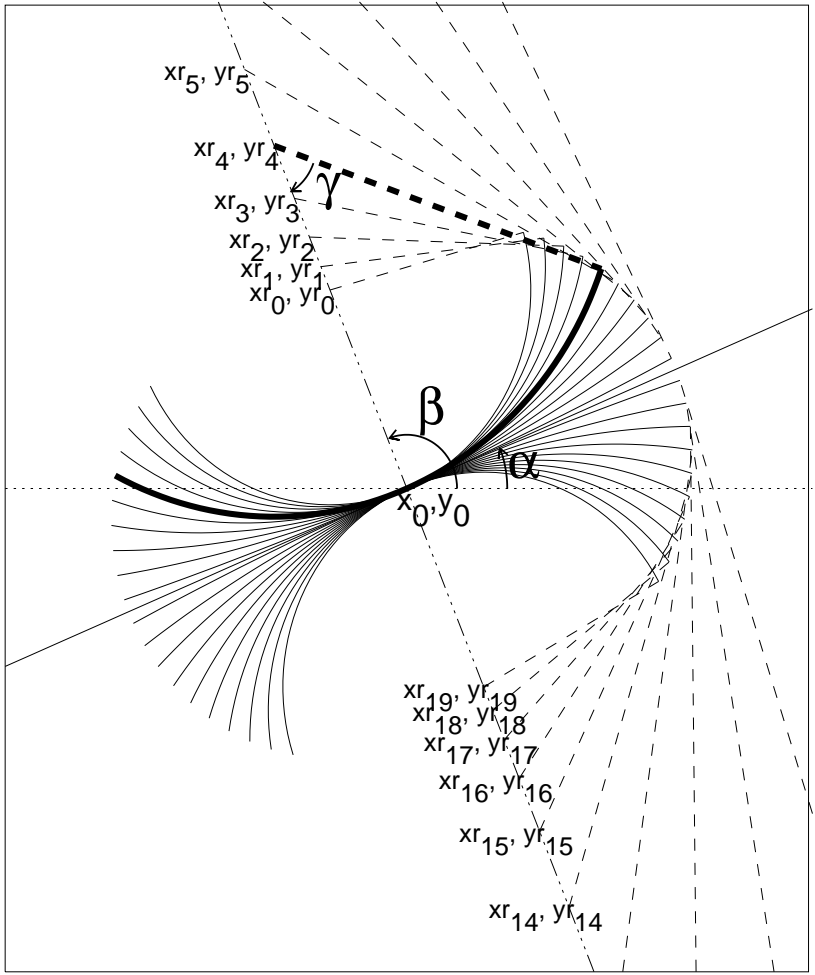}
\caption{The geometry of automated curvi-linear feature tracking
is shown, starting at a local flux maximum location $(x_0,y_0)$,
where the linear direction of the local ridge is measured
(angle $\alpha$) and a set of circular segments within a range
of curvature radii is fitted to the local ridge (thick linestyle).
The locations $(xr_i, yr_i), i=0,...,19$ mark the centers of the
curvature radii (Aschwanden, De Pontieu, and Katrukha 2013a).}
\end{figure}

\begin{figure}
\plotone{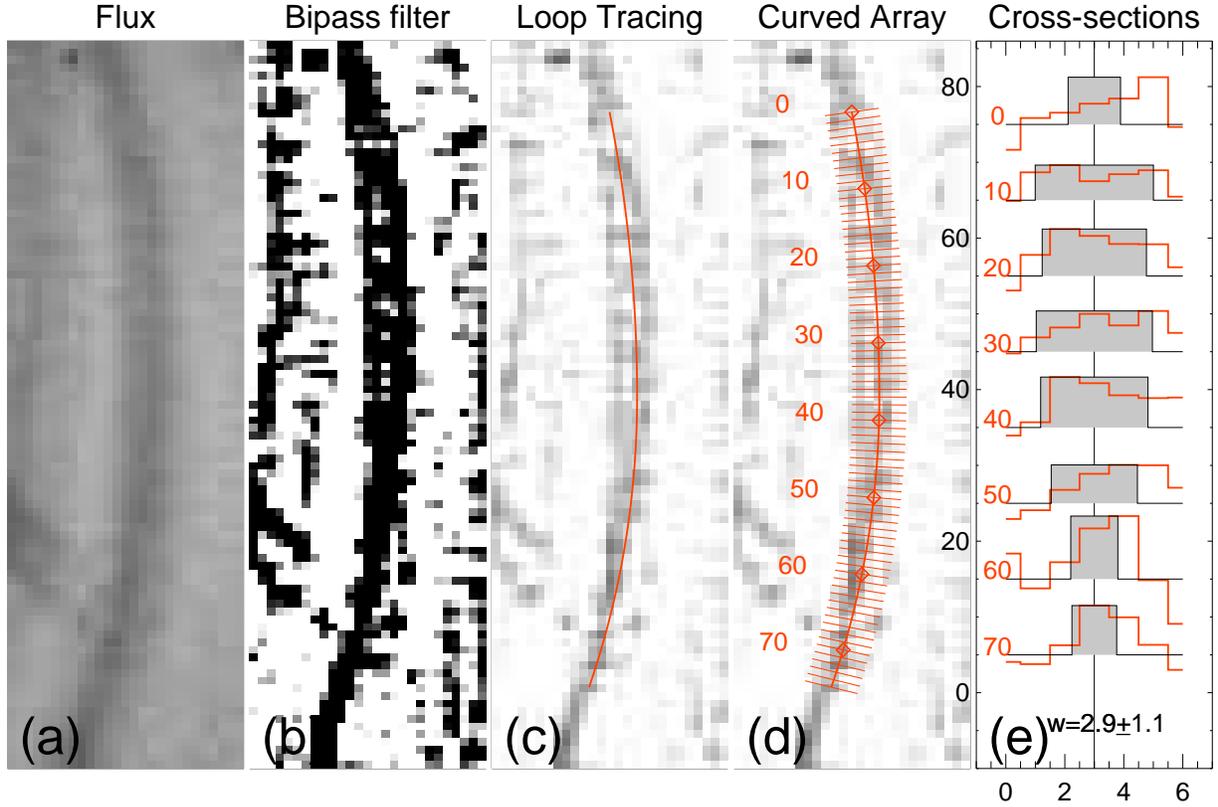}
\caption{Example of loop width measurements along a loop segment:
(a) flux of original
AIA image; (b) bipass-filtered image with highpass filter
of $n_{sm1}=3$ pixels and lowpass filter $n_{sm2}=5$ pixels;
(c) Automated loop tracings above a noise threshold (of 2 standard
deviations of background) with OCCULT-2 code; (d) Curved array 
along a traced loop (red curve) with a width range of 
$n_w=n_{sm1}+4=7$ pixels; (e) Interpolated cross-sectional flux 
profiles (red histograms) in perpendicular direction to the loop axis,  
with equivalent-width measurements (grey rectangles), yielding
a mean width of $w=2.9\pm1.1$ pixels.}
\end{figure}

\begin{figure}
\plotone{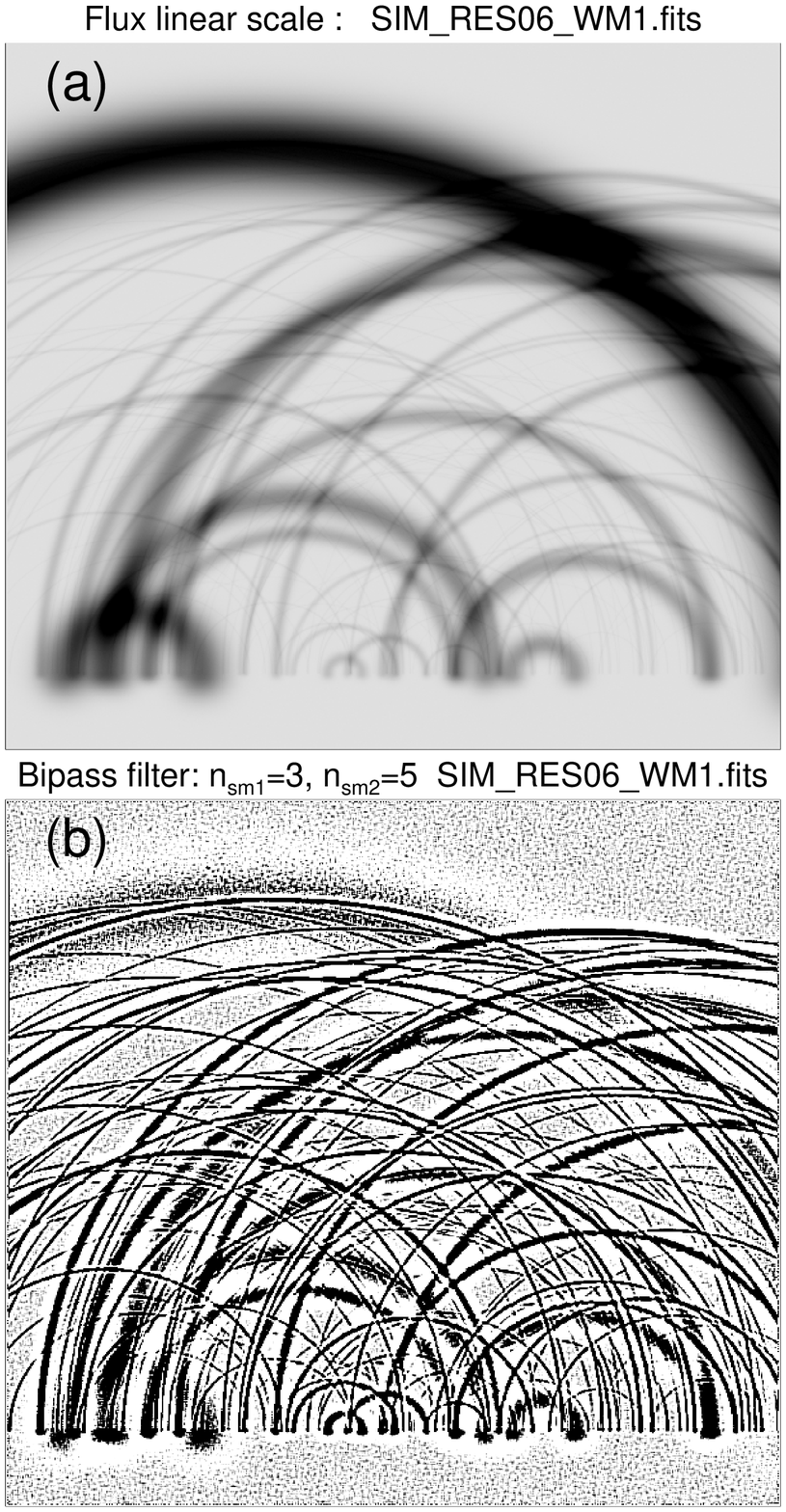}
\caption{(a): Monte-Carlo simulation of loop image with 
128 randomly distributed loops that have a power law distribution 
of loop widths $N(w_{sim}) \propto (w_{sim}+w_0)^{-a}$ with a 
power law slope $a=2.0$ and resolution $w_0=1$ pixel;
(b): Bipass-filtered image with filters $n_{sm1}=3$ and
$n_{sm2}=5$, which filters out structures with widths of 
$w = 4 \pm 1$.}
\end{figure}

\begin{figure}
\plotone{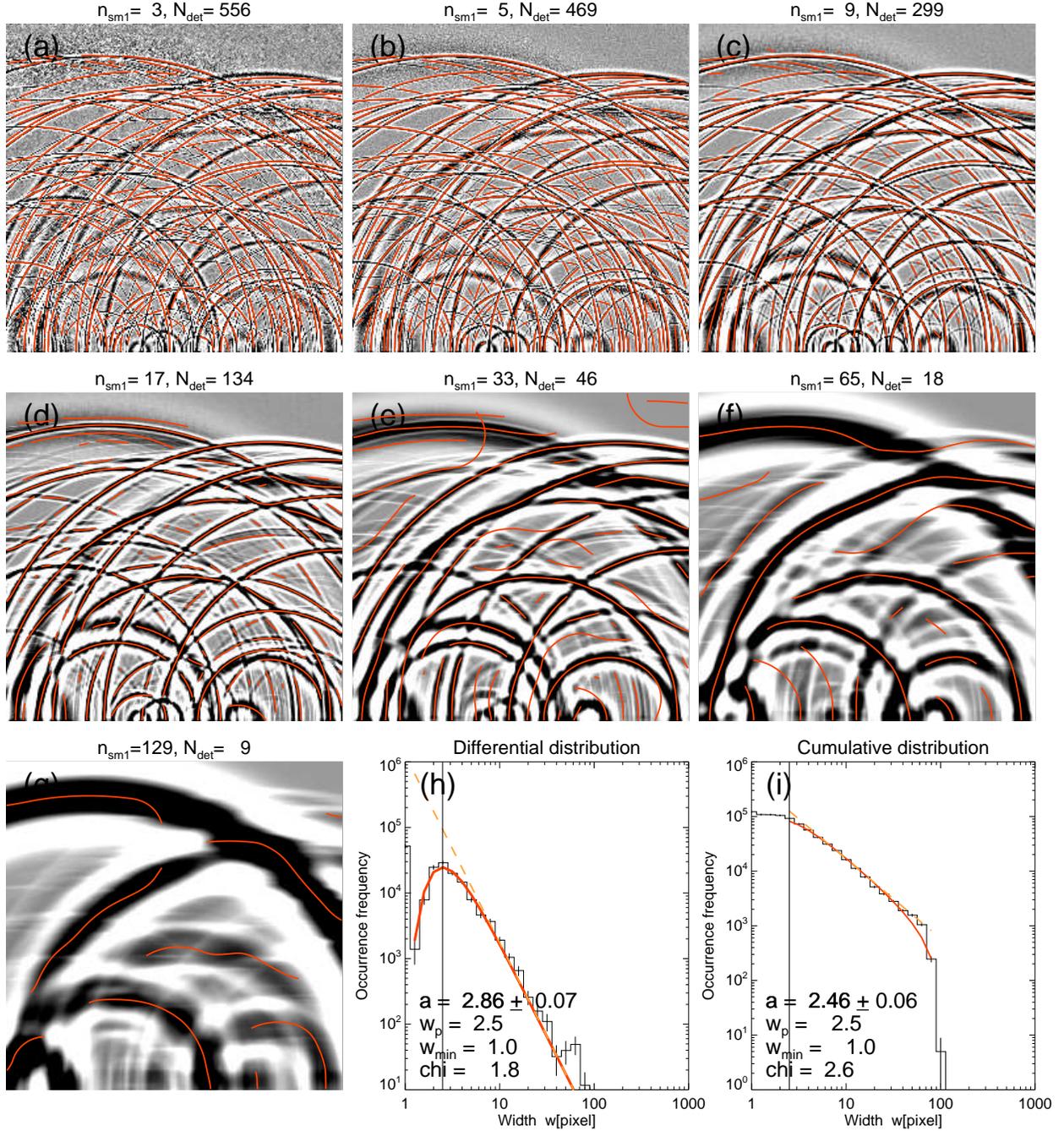}
\caption{Seven bipass-filtered images (grey scale) with filters of
$n_{sm1}=2, 4, 8, 16, 32, 64, 128$ pixels, produced from the original
image shown in Fig.~5a. The automated loop detection is visualized
with red curves in each of the 7 filters. The differential and
cumulative size distributions of the histogrammed loop widths
(histograms with error bars) are fitted with the theoretical
distribution function consisting of a power law with a smooth
cutoff (Fig.~2b), defined in Eq.~(11) (red curves). The power law
slope is indicated with a dashed orange line.}
\end{figure}

\begin{figure}
\plotone{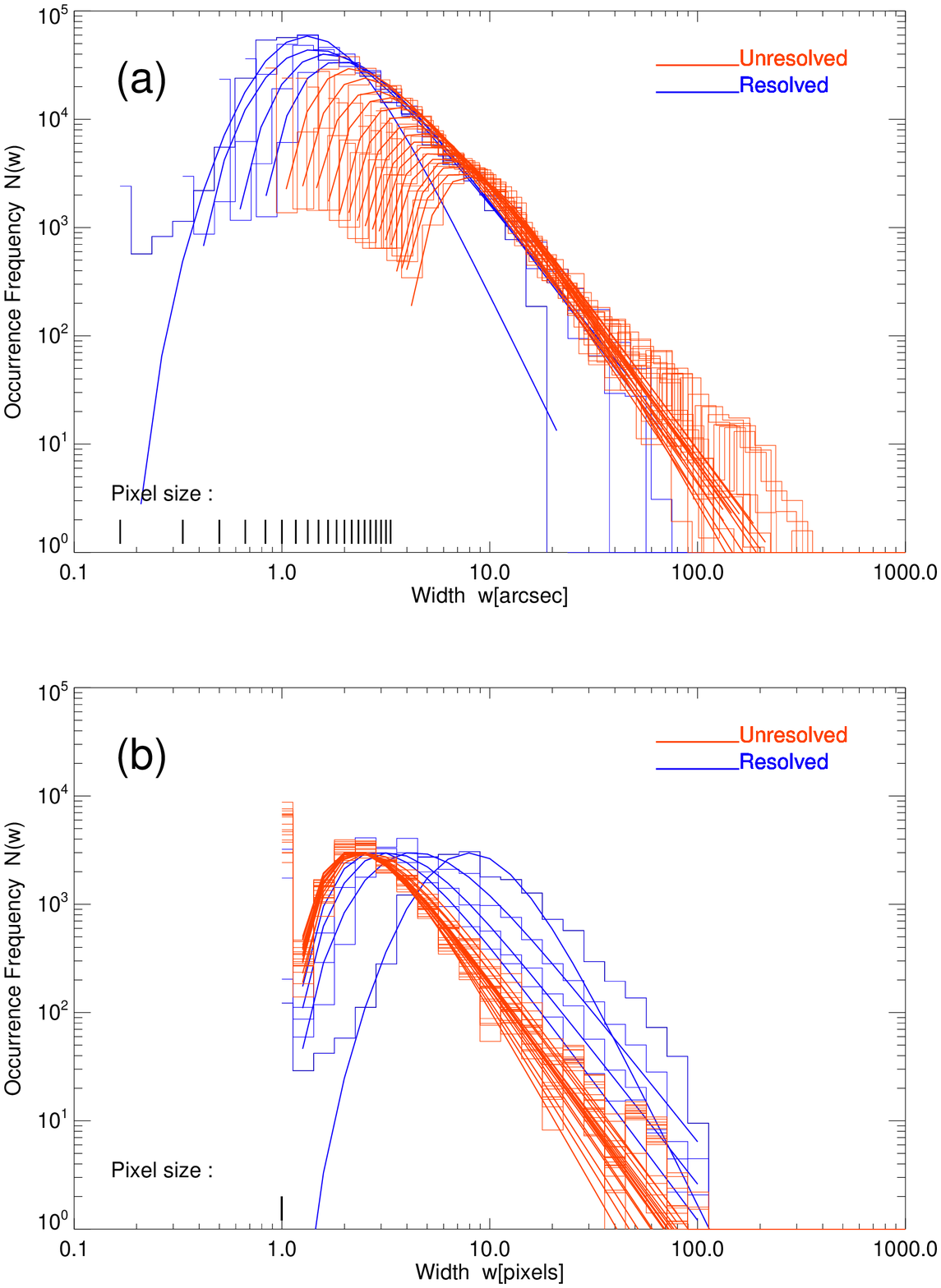}
\caption{(a) A set of 20 size distributions from identical Monte-Carlo 
simulations is shown (with a minimum loop width of $w_{min}=0.6\arcsec$,
but different pixel sizes or spatial resolutions (for
$\Delta x=0.1\arcsec, 0.2\arcsec, ..., 2.0\arcsec$).  
The obtained size distributions are 
rendered as histograms, the analytical fits of the unresolved cases 
($w_{min} < \Delta x$) with red curves, and the resolved cases
($w_{min} > \Delta x$) with blue curves. (b) The fits are normalized 
by their pixel sizes and peak occurrence rates in diagram (b), 
which shows the universal shape (red curves) for the unresolved cases.}
\end{figure}

\begin{figure}
\plotone{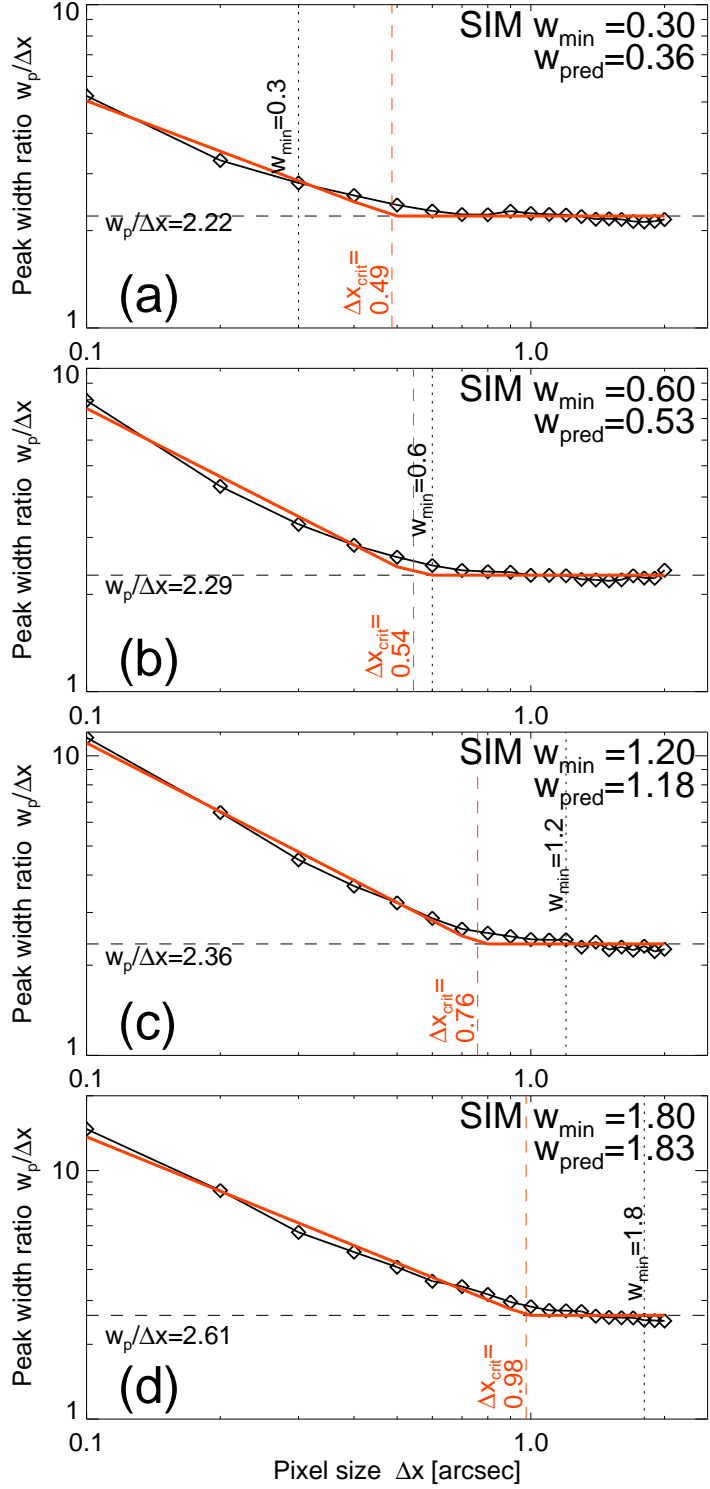}
\caption{The peak widths (normalized by the pixel size) $w_p/\Delta x$, 
are shown as a function of the pixel size $\Delta x$, 
measured from 20 simulated size distributions with pixel sizes of 
$\Delta x=0.1\arcsec,...,2.0\arcsec$, for four sets of numerical
simulations, with 
$w_{min}=0.3\arcsec$ pixel (a), 
$w_{min}=0.6\arcsec$ pixel (b), 
$w_{min}=1.2\arcsec$ pixels (c), and 
$w_{min}=1.8\arcsec$ pixels (d). The simulated values are indicated
with diamonds, and best-fit models are overlaid with red solid curves. 
The simulated minimum loop widths $w_{min}$ are indicated with black 
vertical lines, while the predicted values $w_{pred}$ are shown with 
red (dashed) vertical lines, predicted by the critical value 
$w_p/\Delta x = 3.0$ (horizontal black dashed line).}
\end{figure}

\begin{figure}
\plotone{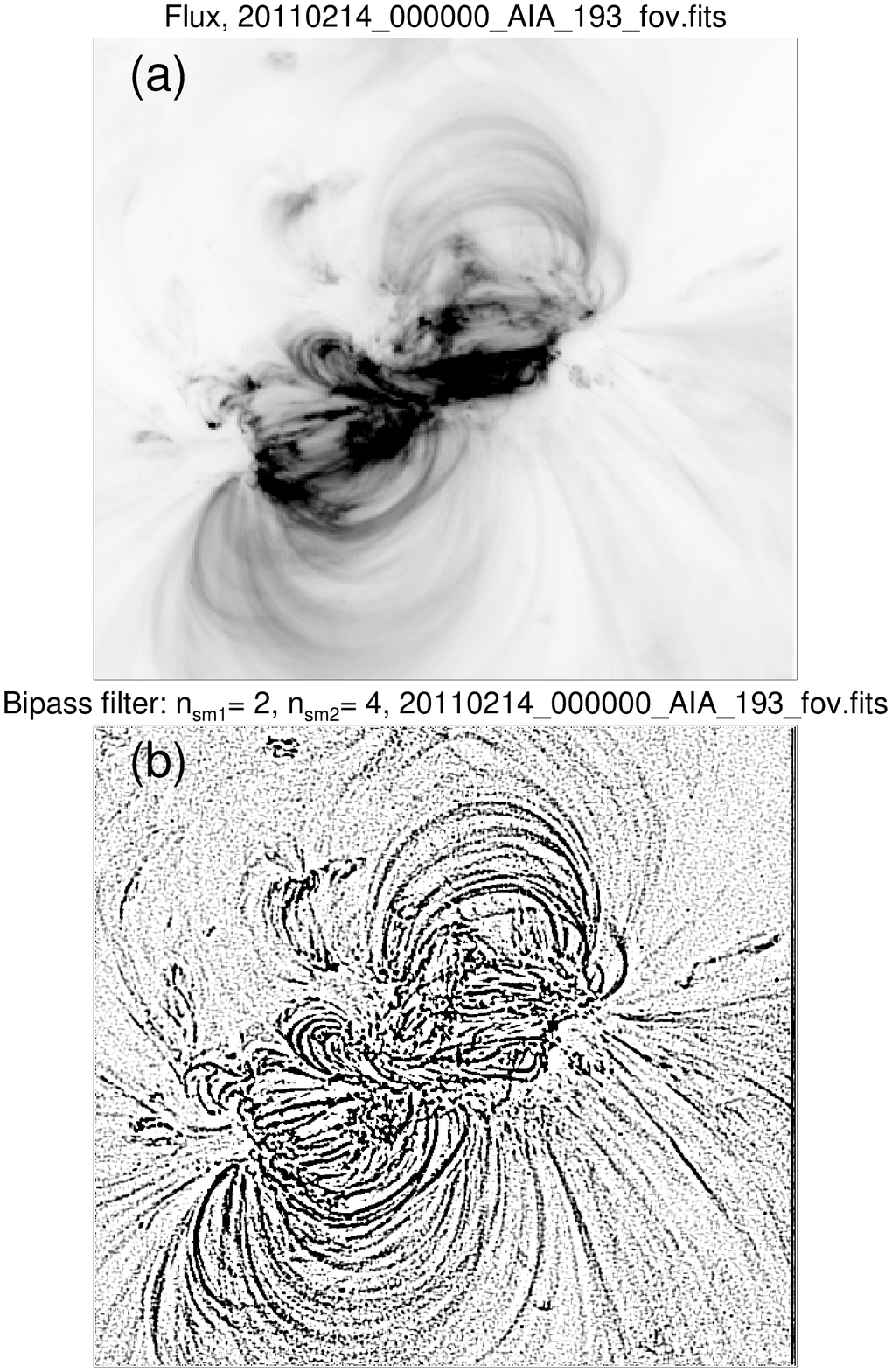}
\caption{(a) EUV image observed with AIA/SDO on 2011-Feb-14, 00:00:00 UT,
at a wavelength of 193 \ang . The grey scale is inverted (black for
bright emission) and the flux is rendered on a linear scale.
(b): Bipass-filtered image with filters $n_{sm1}=1$ and
$n_{sm2}=3$, which filters out structures with widths of 
$w = 3 \pm 1$. The image has a size of $486 \times 486$ and
a linear extent of 211 Mm.}
\end{figure}

\begin{figure}
\plotone{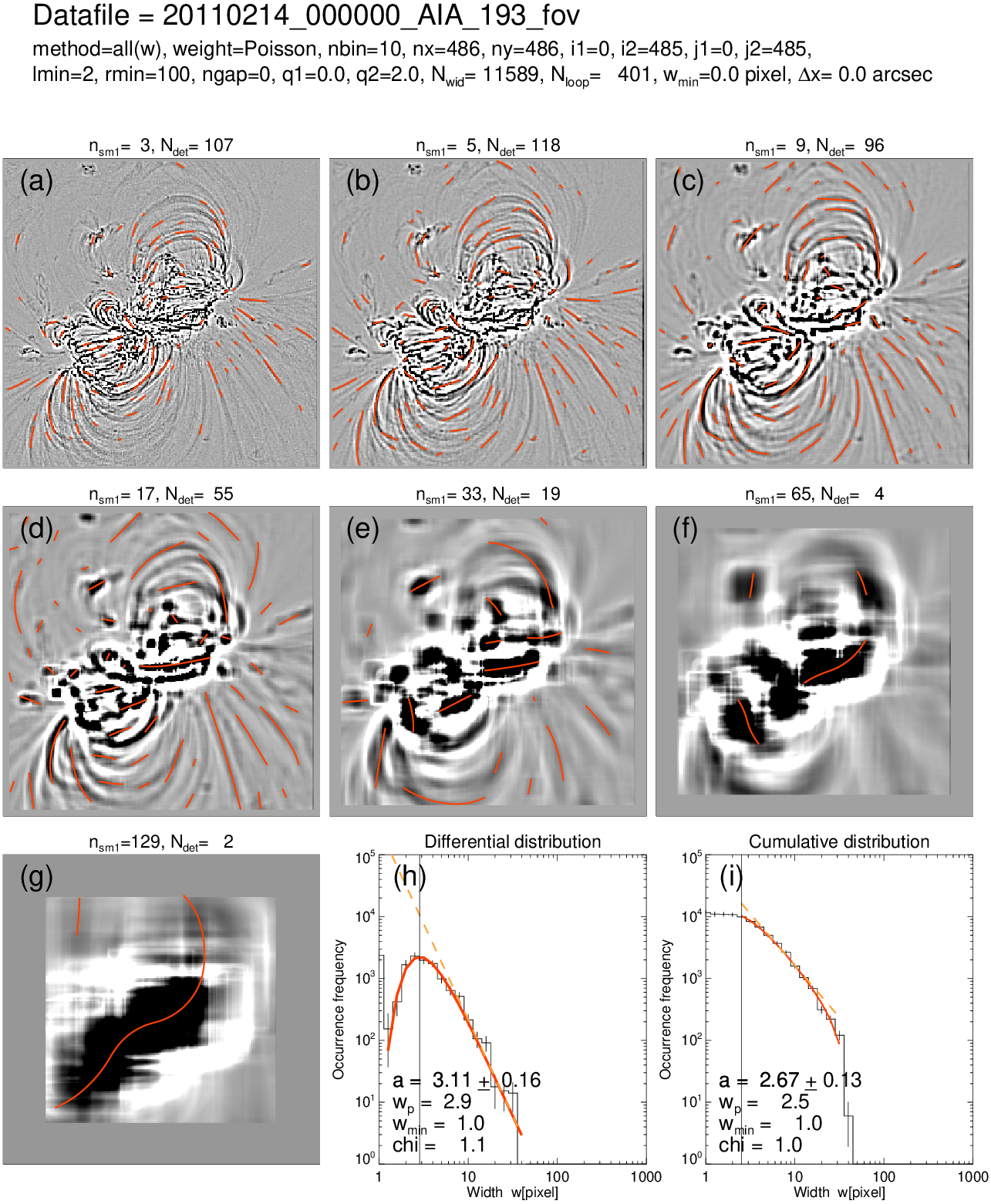}
\caption{Seven bipass-filtered images (grey scale) with filters of
$n_{sm1}=2, 4, 8, 16, 32, 64, 128$ pixels and automated loop tracings
(red curves), obtained from the AIA/SDO image at a wavelength of 
193 \ang\ as shown in Fig.~9. Otherwise similar representation 
as in Fig.~6.}
\end{figure}

\begin{figure}
\plotone{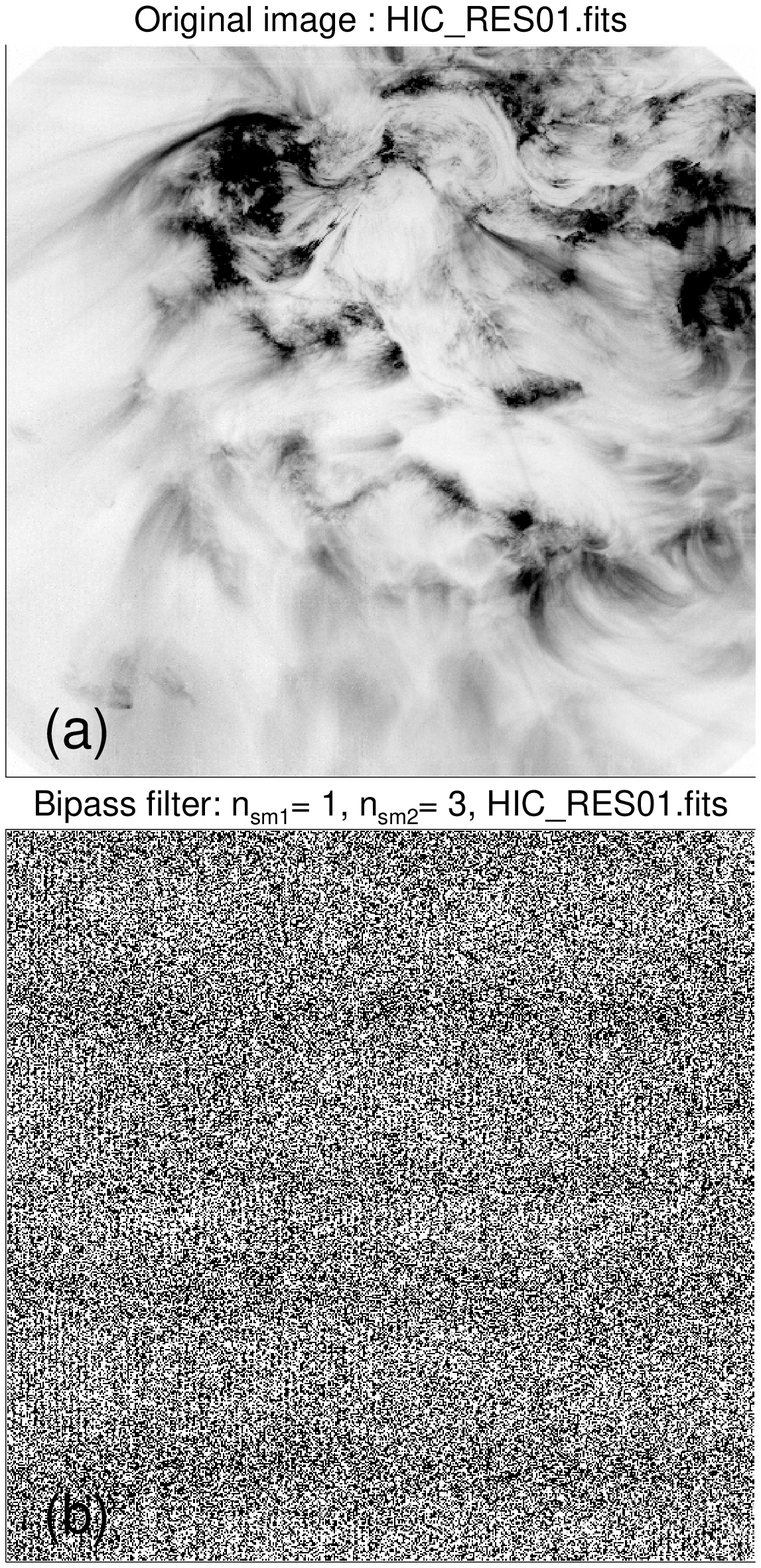}
\caption{(a) EUV image observed with Hi-C on 2012-Jul-11, 18:54:16 UT,
at a wavelength of 193 \ang . The grey scale is inverted (black for
bright emission) and the flux is rendered on a linear scale.
(b): Bipass-filtered image with filters $n_{sm1}=1$ and
$n_{sm2}=3$, which selects structures with a widths of 
$w \approx 140$ km. Note that the filtered image mostly shows
random noise without any coherent loop strand structures.}
\end{figure}

\begin{figure}
\plotone{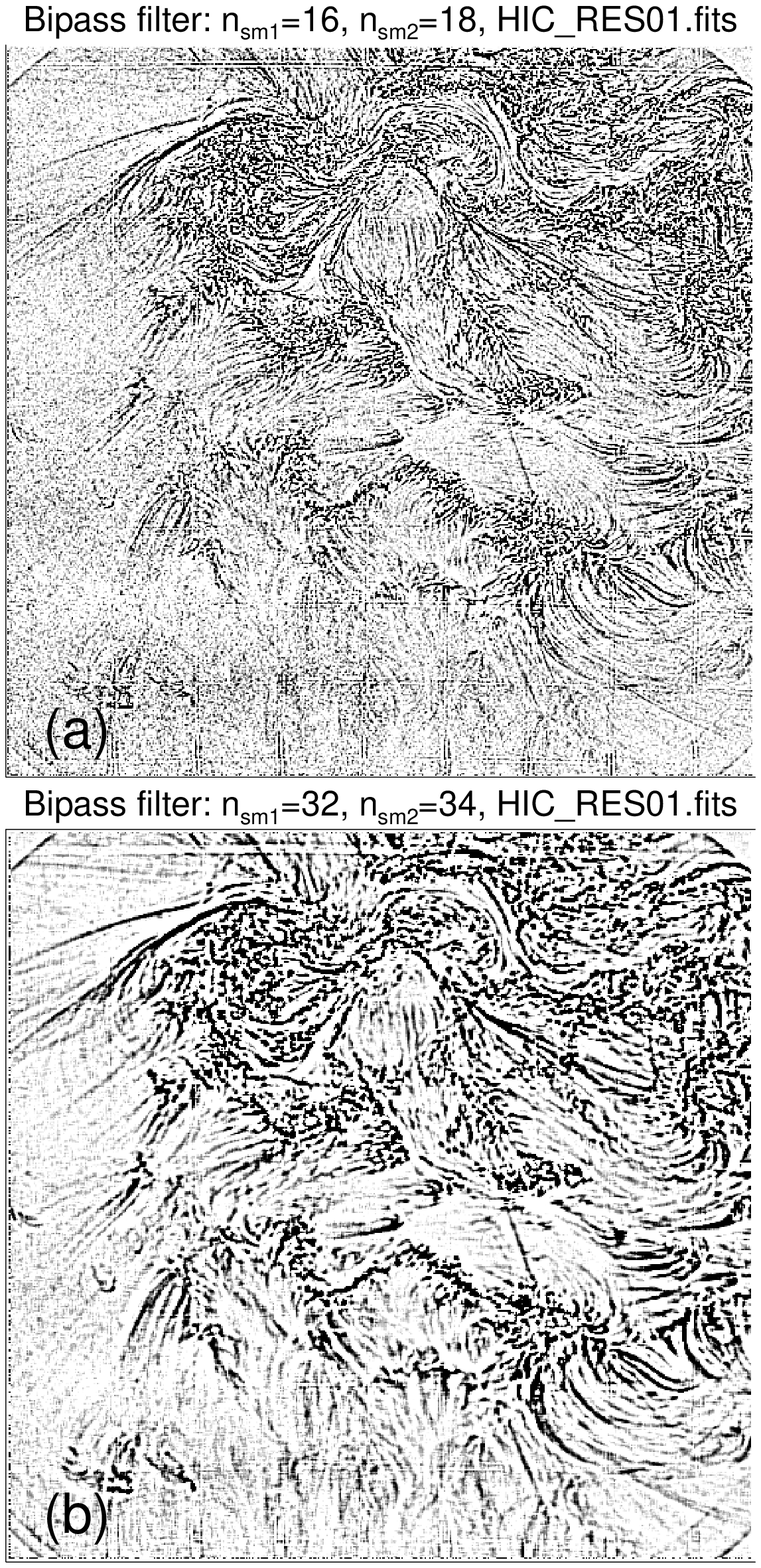}
\caption{The same Hi-C image as shown in Fig.~9a, but filtered
with $n_{sm1}=16$ and $n_{sm2}=18$, selecting structures with
widths of $w \approx 1200$ km (a), and filtered with 
with $n_{sm1}=32$ and $n_{sm2}=34$, selecting structures with
widths of $w \approx 2500$ km (b).}  
\end{figure}

\begin{figure}
\plotone{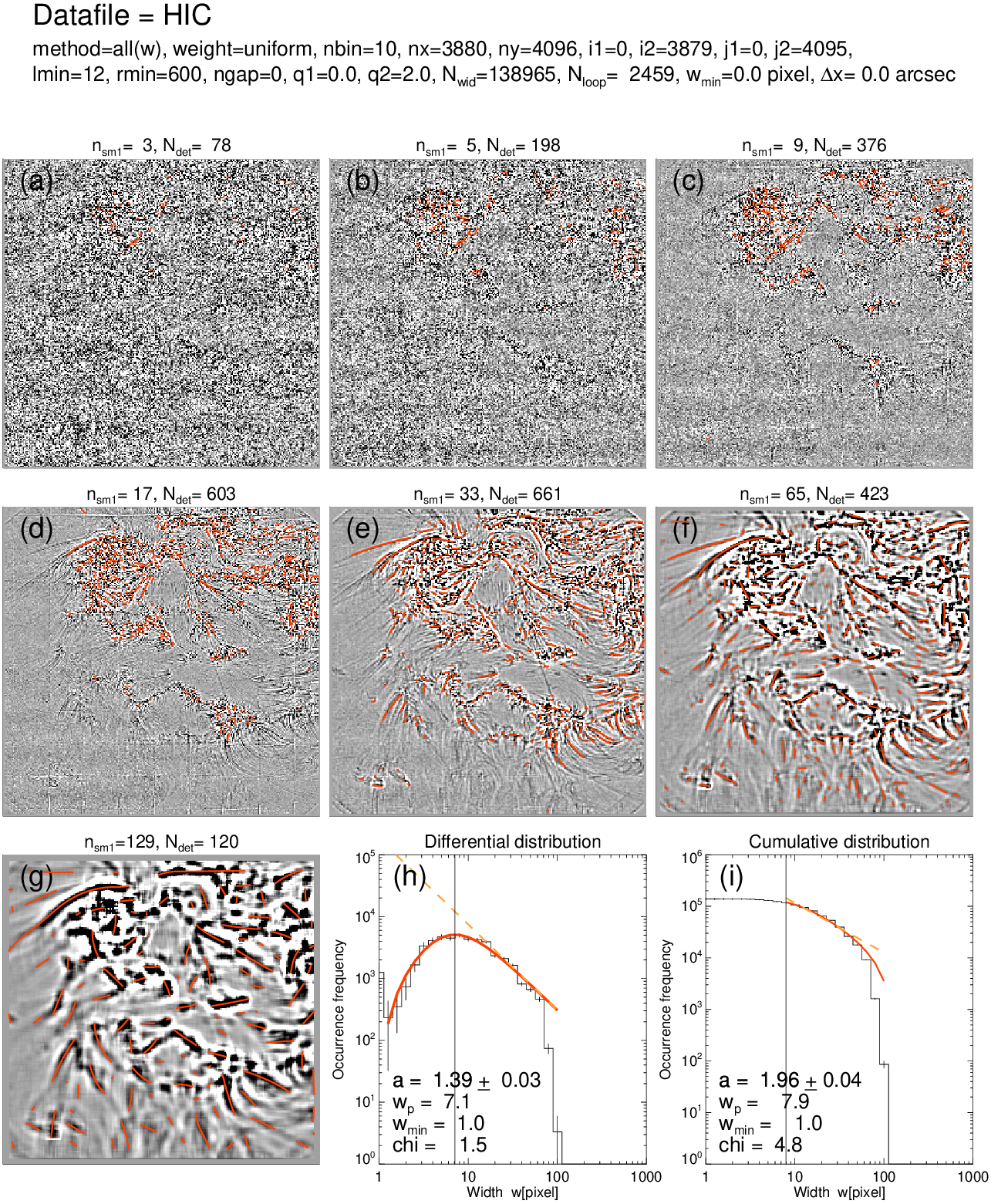}
\caption{Seven bipass-filtered images (grey scale) with filters of
$n_{sm1}=2, 4, 8, 16, 32, 64, 128$ pixels and automated loop tracings
(red curves), obtained from the HiC image at a wavelength of 
193 \ang\ as shown in Fig.~11a. Otherwise similar representation 
as in Fig.~6.}
\end{figure}

\begin{figure}
\plotone{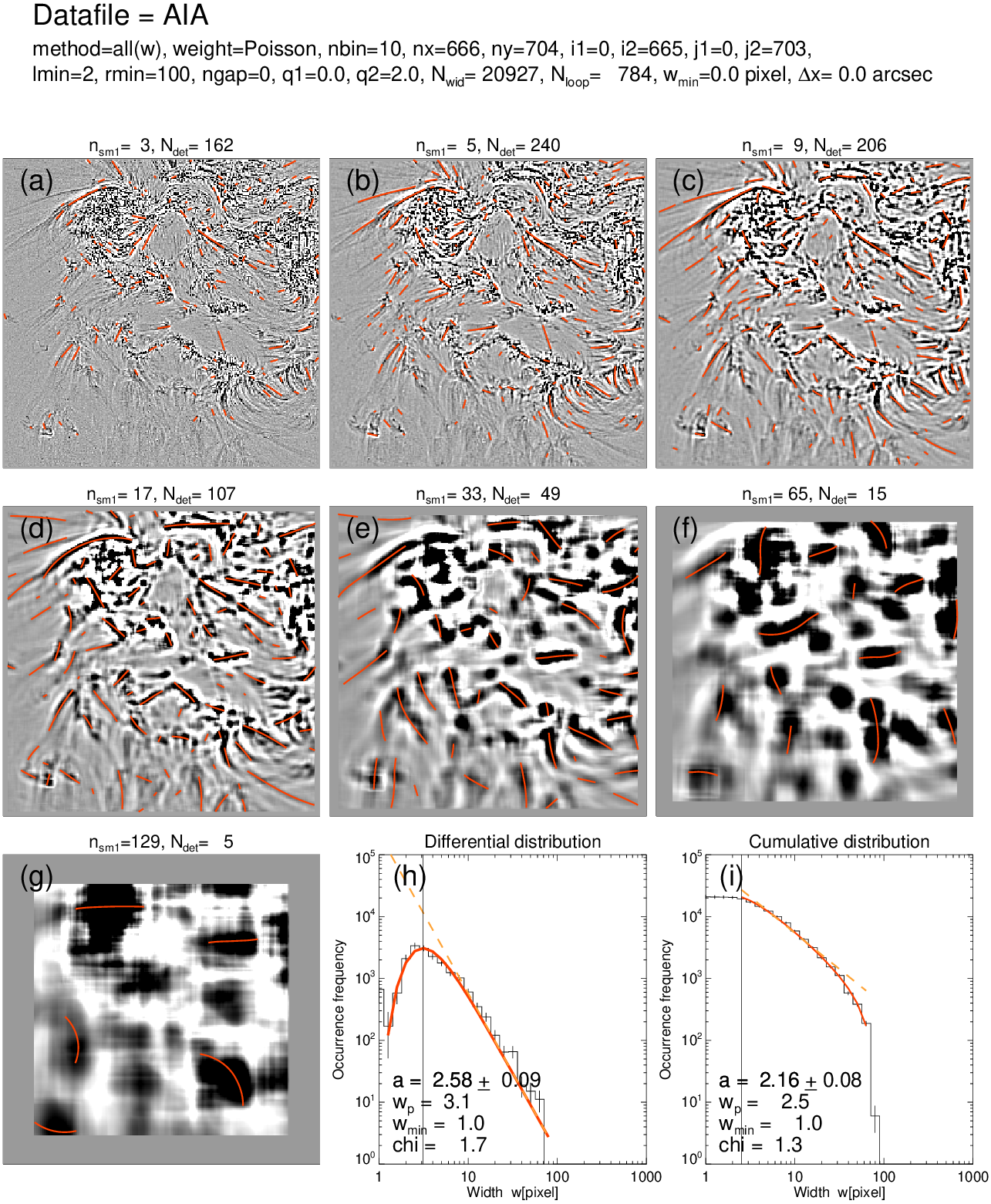}
\caption{Seven bipass-filtered images (grey scale) with filters of
$n_{sm1}=2, 4, 8, 16, 32, 64, 128$ pixels and automated loop tracings
(red curves), obtained from the AIA/SDO image at a wavelength of 
193 \ang . Otherwise similar representation as in Fig.~13.}
\end{figure}

\begin{figure}
\plotone{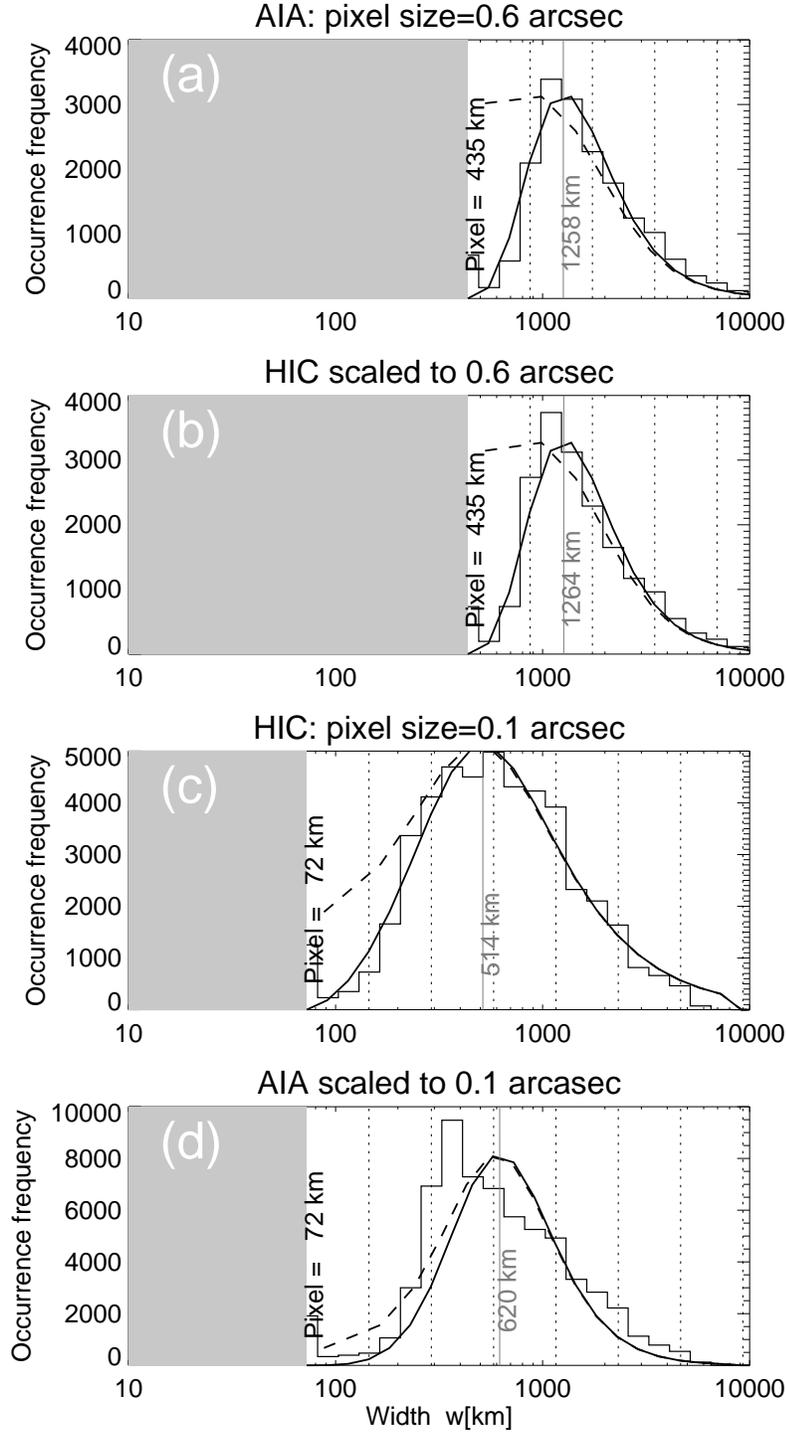}
\caption{Comparison of width size distributions $N(w)$ (histograms)
with fits of the model of Eq.~11 (solid curve), and corrected for
finite resolutions (dashed curve), obtained with AIA (a), HiC scaled 
to AIA resolution (b), HiC (c), and AIA scaled to HiC resolution (d). 
The grey areas indicate width ranges where no structures can be resolved.
The vertical dotted lines indicate the discrete width filters
$n_{sm1}=2, 4, 8, ..., 128$.}
\end{figure}

\begin{figure}
\plotone{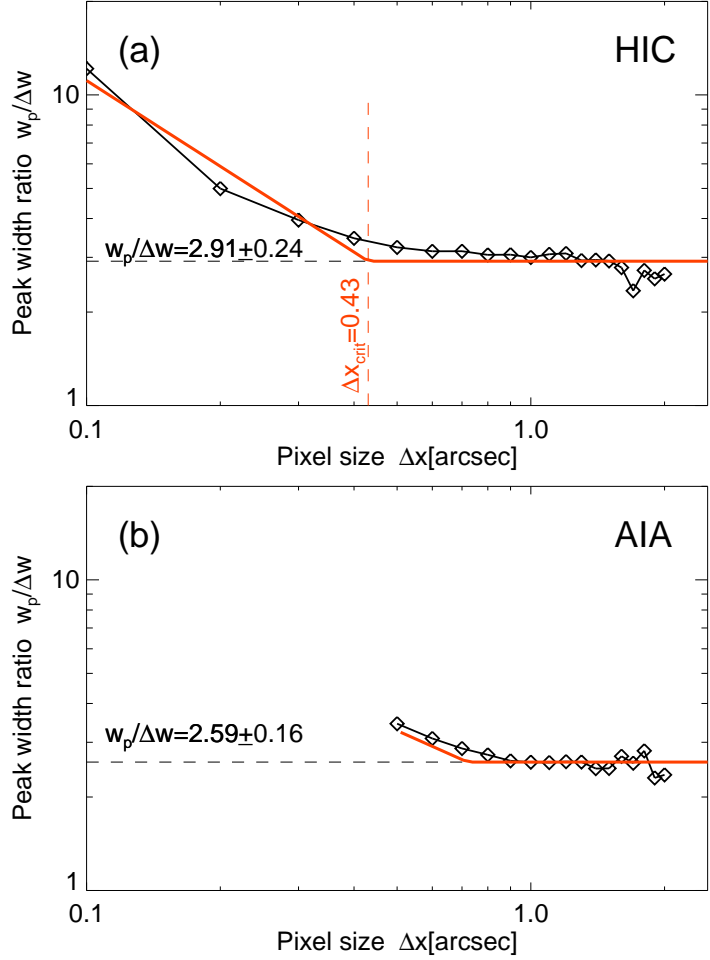}
\caption{The peak width $w_p$ (of the width size distribution) is shown
as a function of the pixel size $\Delta x \ge 0.1\arcsec$ for the
analyzed HI-C image (a), and for $\Delta x \ge 0.6\arcsec$ for the
simultaneously observed AIA image (b). At $w_p/\Delta x=3.0$, a minimum 
width of $\Delta x=0.77\arcsec \approx 550$ km is found
for the Hi-C image, below which loop structures are over-resolved
with Hi-C. AIA resolves the structures at $\Delta x \lapprox 0.58\arcsec$
= 420 km marginally.}
\end{figure}
\clearpage

\begin{figure}
\plotone{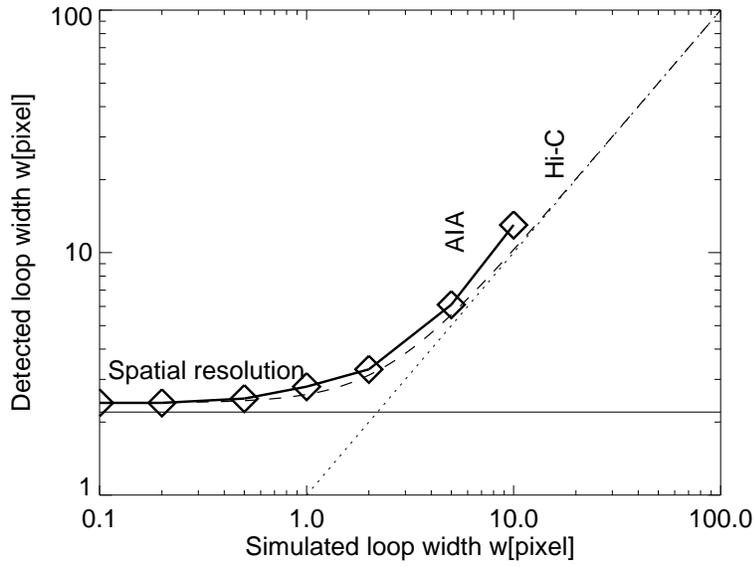}
\caption{Detected (or observed) loop width ratio $w_p/w_{min}$ 
as a function of simulated (or true) loop widths $w_{true}$
are shown, simulated with different true loop widths $w_{true}$. 
The diamonds represent the ratios of the peak widths $w_p$ to 
the minimum widths $w_{min}$ of seven simulated width distributions. 
The horizontal line indicates the limit of spatially unresolved structures,
which yields an asymptotic value of $w_p/w_{min} \approx 2.5\arcsec$. 
The dotted line indicates equivalence between simulated and detected 
loop widths.  The dashed line indicates the relationship 
$w_{det}=\sqrt{w_{true}^2+w_{min}^2}$ (Eq.~4).}
\end{figure}
\clearpage

\begin{figure}
\plotone{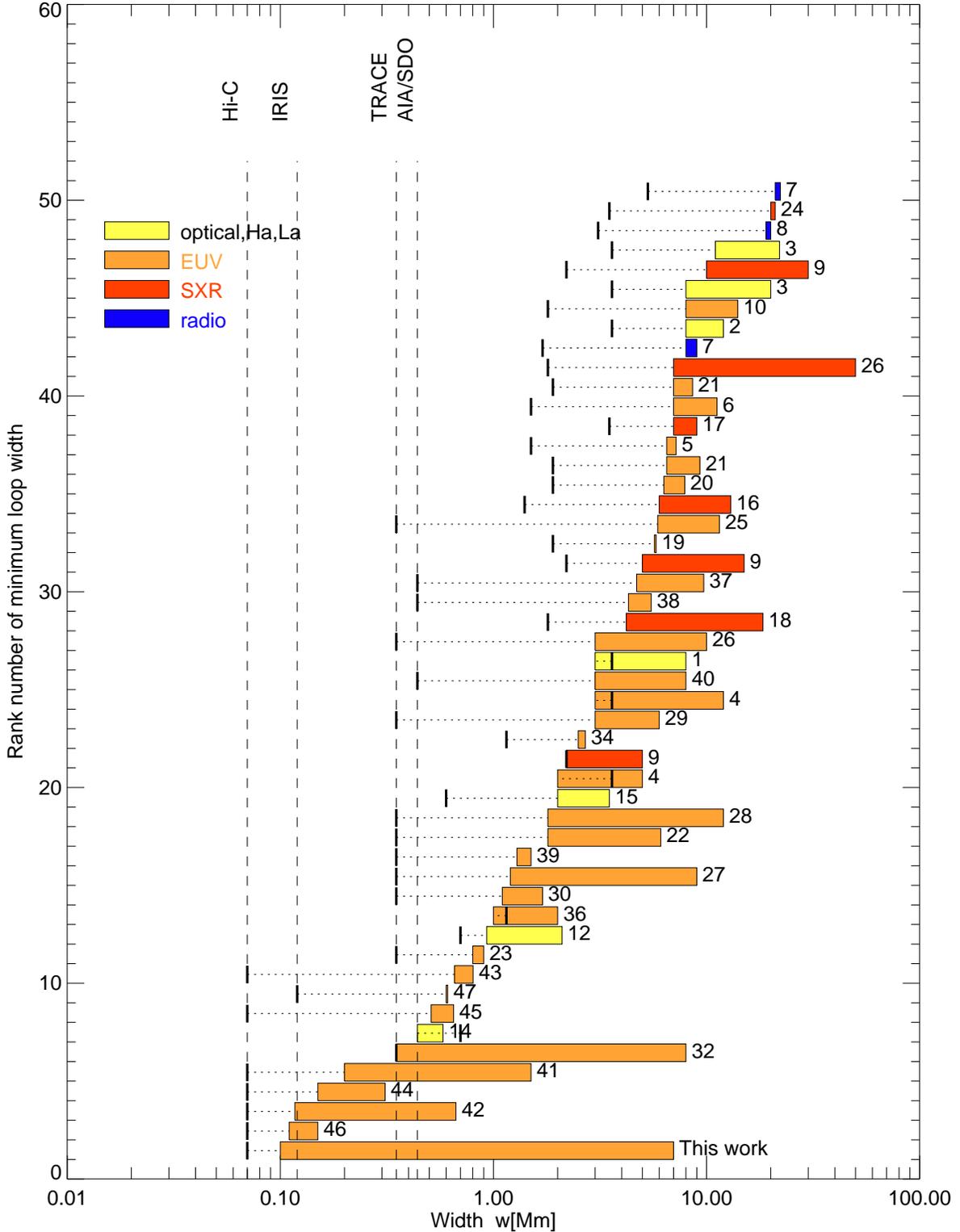}
\caption{Compilation of loop width measurements from literature during
1963-2016: The measured ranges are represented by a block, colored by
wavelength regimes (yellow=optical, H$\alpha$, L$\alpha$; orange=EUV,
red=SXR, and blue=radio), labeled with the reference number given in Table 3,
and sorted by the increasing minimum width on the y-axis.
The instrumental pixel size (or resolution if pixel size is not known) is 
indicated with a black thick bar for each measurement. Automatically
detected structures with widths of $w \gapprox 2$ Mm in this study
are suspected to represent more complex structures than traditional loops.}
\end{figure}
\clearpage

\end{document}